%% file: rcm-ewens-results.tex
\documentclass[10pt,oneside]{article}
\usepackage{verbatim}
\usepackage{fullpage}

\usepackage{amssymb,amsmath,amsthm}
\usepackage[mathscr]{eucal}

\usepackage{makeidx}
\numberwithin{equation}{section}
\input{isomath}
\input{mathenv}

\input{syms}

\usepackage{graphicx}
\usepackage{psfrag}
\usepackage{afterpage}
\usepackage{subfigure}
\usepackage{floatflt}

\makeindex

\usepackage{hyperref}
\usepackage{ifpdf}

\ifpdf
\fi

\begin{document}
\title{Shift in critical temperature for random spatial permutations with cycle weights}
\author{John Kerl}
\date{\today}

\maketitle

\begin{abstract}
\input{resultsabs.tex}
\end{abstract}


\vspace*{20pt}

\input{resultsbody.tex}

\section*{}
\addcontentsline{toc}{section}{References}
\input{resultsbib.tex}


\end{document}

%% file: isomath.tex
\newcommand*{\bbb}[1]{\mathbb{#1}}

\newcommand{\bR}{\bbb{R}}

\newcommand{\R}{\bR}
\newcommand{\Z}{\bZ}

\usepackage{amsbsy}



%% file: mathenv.tex
\usepackage{ifthen}

\theoremstyle{plain}

\newtheorem{lem}[equation]{Lemma}
\newtheorem{prop}[equation]{Proposition}

\newtheorem*{thm*}{Theorem}
\newtheorem*{cor*}{Corollary}
\newtheorem*{lem*}{Lemma}
\newtheorem*{prop*}{Proposition}
\newtheorem*{ax*}{Axiom}

\theoremstyle{definition}
\newtheorem{defn}[equation]{Definition}

\newtheorem*{defn*}{Definition}
\newtheorem*{problem*}{Problem}
\newtheorem{ex}[equation]{Example}

\newtheorem*{ex*}{Example}
\newtheorem*{exs*}{Examples}

\newtheorem*{rem*}{Remark}

\newtheorem*{aside*}{Aside}

\newtheorem*{intuition*}{Intuition}
\newtheorem{notn}[equation]{Notation}
\newtheorem*{notn*}{Notation}

\newtheorem*{conv*}{Convention}

\newtheorem*{interp*}{Interpretation}

\newtheorem*{mnem*}{Mnemonic}

\newtheorem*{exer*}{Exercise}

\newtheorem*{conjecture*}{Conjecture}

\newcommand{\beginex}{\begin{ex}$\rhd$ }
\newcommand{\mendex}{\hfill$\lhd$\end{ex}}

\newcounter{ppart}



%% file: syms.tex
\newcommand*{\emphidx}[1]{\index{#1}\emph{#1}}
\newcommand*{\plainidx}[1]{\index{#1}#1}

\newcommand{\ellmax}{\ell_{\textrm{max}}}
\newcommand{\fM}{f_{\textrm{max}}}

\def\veps{\varepsilon}

\def\bbE{\mathbb{E}}

\def\nuhat{\hat{\nu}}
\def\rhohat{\hat{\rho}}

\def\torusdiff{\vecd_\Lambda}

\def\Z{\mathbb{Z}}

\newcommand*{\pig}[1]{\langle #1 \rangle}

\newcommand{\SN}{\mathcal{S}_N}

\newcommand{\Var}{\mathrm{Var}}

\newcommand{\Cov}{\mathrm{Cov}}

\def\veca{\mathbf{a}}
\def\vecb{\mathbf{b}}

\def\vecd{\mathbf{d}}

\def\veck{\mathbf{k}}
\def\vecm{\mathbf{m}}
\def\vecn{\mathbf{n}}

\def\vecx{\mathbf{x}}
\def\vecy{\mathbf{y}}
\def\vecz{\mathbf{z}}

\def\vecM{\mathbf{M}}
\def\vecP{\mathbf{P}}

\def\vecW{\mathbf{W}}

\def\vecX{\mathbf{X}}

\def\vecxpii{\mathbf{x}_{\pi(i)}}

\def\pix{\pi(\vecx)}
\def\piy{\pi(\vecy)}

\def\edh{e^{-\Delta H}}

\def\maybeeq{\stackrel{?}{=}}

\newcommand{\tand}{\textrm{and}}

\newcommand{\qor}{\quad\textrm{or}\quad}

\newcommand*{\colvecthree}[3]{\left(\begin{array}{r} #1 \\ #2 \\ #3\end{array}\right)}

\newcommand*{\pcmatrix}[1]{
	\left(\begin{array}{cccccccccccc}
		#1
	\end{array}\right)
}



%% file: resultsabs.tex
We examine a phase transition in a model of random spatial permutations which
originates in a study of the interacting Bose gas.  Permutations are weighted
according to point positions; the low-temperature onset of the appearance of
arbitrarily long cycles is connected to the phase transition of Bose-Einstein
condensates.  In our simplified model, point positions are held fixed on the
fully occupied cubic lattice and interactions are expressed as Ewens-type
weights on cycle lengths of permutations.  The critical temperature of the
transition to long cycles depends on an interaction-strength parameter
$\alpha$.  For weak interactions, the shift in critical temperature is expected
to be linear in $\alpha$ with constant of linearity $c$.  Using Markov chain
Monte Carlo methods and finite-size scaling, we find $c = 0.618 \pm 0.086$.
This finding matches a similar analytical result of Ueltschi and Betz.  We also
examine the mean longest cycle length as a fraction of the number of sites in
long cycles, recovering an earlier result of Shepp and Lloyd for non-spatial
permutations.

%% file: resultsbody.tex
\section{Introduction}
\label{sec:intro}

The model of random spatial permutations arises in the study of the Bose gas.
Its history includes Bose-Einstein, Feynman \cite{bib:feynman}, Penrose-Onsager
\cite{bib:penons}, S\"ut\H{o} \cite{bib:suto1,bib:suto2}, and
Betz-Ueltschi-Gandolfo-Ruiz \cite{bib:gru,bib:qmath,bib:bu07,bib:bu08}.  Such
random permutations arise physically when one symmetrizes the $N$-boson
Hamiltonian with pair interactions, then applies a multi-particle Feynman-Kac
formula and a cluster expansion \cite{bib:bu07,bib:bu08}.  Specifically, given
points $\vecx_1, \ldots, \vecx_N$ in the box $[0,L]^3$ and temperature $T$,
permutations $\pi$ are given probability weights proportional to the Gibbs
factor $e^{-H(\pi)}$ where
\begin{align}
\label{eqn:intro_H}
	H(\pi) &= \frac{T}{4} \sum_{i=1}^N \|\vecx_i-\vecx_{\pi(i)}\|_\Lambda^2
\end{align}
for the non-interacting case.
The notation $\|\cdot\|_\Lambda$ indicates the natural
distance on the 3-torus:
\begin{align}
\label{eqn:prb_model_torus_diff}
	\|\vecx - \vecy\|_\Lambda &=
		\min_{\vecn\in \Z^3} \{\|\vecx - \vecy + L \vecn\|\}
\end{align}
(The sum in equation \eqref{eqn:intro_H} is scaled by temperature rather than
reciprocal temperature.  This surprising feature is opposite that of many
models in statistical mechanics.) These energy terms involve lengths of
permutation jumps; additional interaction terms take the form
\begin{align}
\label{eqn:intro_V}
	\sum_{i<j} V(\vecx_i, \vecx_{\pi(i)}, \vecx_j, \vecx_{\pi(j)}),
\end{align}
i.e. permutation jumps from sites $\vecx_i$ and $\vecx_j$ interact pairwise.
In the above-cited papers of Betz and Ueltschi, these may be approximated and
rearranged such that one obtains interaction terms of the form
\begin{align}
\label{eqn:intro_ell}
	\sum_{\ell=1}^N \alpha_\ell r_\ell(\pi)
\end{align}
where $r_\ell(\pi)$ counts the number of $\ell$-cycles of the permutation
$\pi$, and the coefficients $\alpha_\ell$ are \emph{cycle weights}.

When the Bose gas is cooled below a critical temperature $T_c$, there is a
phase transition: a macroscopic fraction of the bosons are found in the ground
state of the external potential, and such particles are said to participate in
a Bose-Einstein condensate.  In the permutation representation, this transition
manifests itself as the onset of long permutation cycles.  Bose suggested the
statistics carrying his name for describing the gas of photons; Einstein
developed the notion of what we now call Bose-Einstein condensation, and
computed the critical temperature for the non-interacting Bose gas.  The
critical temperature for liquid helium, where interparticle interactions are
strong, is lower than would be expected \cite{bib:baymetal} for non-interacting
atoms of the same density.  For weakly interacting systems, however, an
emerging consensus is that interactions \emph{increase} the critical
temperature.  See \cite{bib:baymetal,bib:su09} for surveys.  Concretely, for
interactions parameterized by some $\alpha$, one defines
\begin{align*}
	\Delta T_c(\alpha) &= \frac{T_c(\alpha)-T_c(0)}{T_c(0)}.
\end{align*}
It is well accepted that
\begin{align*}
	\lim_{\alpha\to 0} \Delta T_c(\alpha) = c \rho^{1/3}\alpha
\end{align*}
where $\rho$ is the particle density, i.e. that for small $\alpha$ the shift in
critical temperature is linear in $\alpha$.  What is more contentious, as
enumerated in the surveys cited above, is the value of the constant $c$.

The interaction terms (equation \eqref{eqn:intro_V}) for the permutation
representation of the Bose gas are difficult to compute.  Moreover, it is
interesting to consider the model of random spatial permutations for its own
sake.  In \cite{bib:gru}, a simulational approach is taken for points held
fixed on the fully occupied unit lattice in the non-interacting case.  In
the papers \cite{bib:qmath,bib:bu07}, Betz and Ueltschi examine the Bose-gas
permutation weights with point positions allowed to vary on the continuum; an
exact expression for the critical temperature is stated and proved for a
simplified interaction model in which only two-cycles interact.  That is,
interactions are of the form of \ref{eqn:intro_ell} with $\alpha_2 = \alpha$,
where $\alpha$ is related to a hard-core scattering length, and the remaining
cycle weights are zero.  In \cite{bib:bu08}, this approach is extended to a
model in which all the $\alpha_\ell$'s may vary, but with the hypothesis that
$\alpha_\ell$ goes to zero faster than $1/\log(\ell)$.  In this paper, we take
a simulational approach to points on the fully occupied unit lattice,
with cycle weights constant in $\ell$ --- removing the decaying-cycle-weight
hypothesis.  The shift in critical temperature is nonetheless found to match
that predicted by Betz and Ueltschi.

An outline of the paper is as follows.  Section \ref{sec:background} provides
background necessary to understand the results of the paper:  the probability
model is defined in section \ref{subsec:prb_model}; qualitative and
quantitative behavior of long cycles are discussed in sections
\ref{subsec:qual_long_cycles} and \ref{subsec:ordps}, respectively.  Known
results and conjectures are listed in section \ref{subsec:conj}.  In section
\ref{sec:simulational_methods}, the simulational methods are presented.  The
\emph{swap-only} and \emph{swap-and-reverse} algorithms generate simulational
data; these algorithms are proved correct in sections
\ref{subsec:SO_algorithm} through \ref{subsec:SO_correctness}.  The
finite-size-scaling method, which reduces the raw simulational data, is
summarized in section \ref{subsec:fss}.  Section \ref{sec:results} presents
the data and its analysis in full detail:  estimation of critical exponents and
critical temperature in sections \ref{subsec:rho_hat_L} through
\ref{subsec:crossing}, verification of the finite-size-scaling hypothesis in
section \ref{subsec:collapse}, and final results in sections
\ref{subsec:Delta_Tc_and_const} through \ref{subsec:conclusions}.

\section{The model of random spatial permutations}
\label{sec:background}

Here we review concepts from \cite{bib:bu07,bib:bu08}, fixing notation and
intuition to be used in the rest of the paper.

\subsection{The probability model}
\label{subsec:prb_model}

The state space is $\Omega_{\Lambda,N} = \Lambda^N \times \SN$, where
$\Lambda=[0,L]^3$ with periodic boundary conditions; point positions are $\vecX
= (\vecx_1, \ldots, \vecx_N)$ for $\vecx_1, \ldots, \vecx_N \in \Lambda$.  The
Hamiltonian takes one of two forms.  In the first, relevant to the Bose gas, we
have
\begin{align}
\label{eqn:results_Bose_H}
	H(\vecX,\pi) &= \frac{T}{4} \sum_{i=1}^N \|\vecx_i-\vecxpii\|_\Lambda^2
		+ \sum_{i<j} V(\vecx_{i}, \vecx_{\pi(i)}, \vecx_{j}, \vecx_{\pi(j)})
\end{align}
where $T=1/\beta$ and the $V$ terms are interactions between permutation jumps.
(The temperature scale factor $T/4$, not $\beta/4$, is surprising but correct
for the Bose-gas derivation of the Hamiltonian.) In the second form of the
Hamiltonian, considered in this paper, we use interactions which are dependent
solely on cycle length:
\begin{align}
\label{eqn:results_RCM_H}
	H(\vecX,\pi) &= \frac{T}{4} \sum_{i=1}^N \|\vecx_i-\vecxpii\|_\Lambda^2
		+ \sum_{\ell=1}^N \alpha_\ell r_\ell(\pi),
\end{align}
where $r_\ell(\pi)$ is the number of $\ell$-cycles in $\pi$ and the
$\alpha_\ell$'s are free parameters, called \emph{cycle weights}.  One
ultimately hopes to choose the $\alpha_\ell$'s appropriately for the Bose gas;
even if not, the model is well-defined and of its own interest.

Different choices of $\alpha_\ell$ result in different models: The
\emph{non-interacting model} \cite{bib:gru} has $\alpha_\ell \equiv 0$.  The
\emph{two-cycle model} \cite{bib:bu07,bib:qmath}, has $\alpha_2=\alpha$ and
other cycle weights are zero.  The \emph{general-cycle model} has no
restrictions on $\alpha_\ell$.  In \cite{bib:bu08}, the \emph{decaying
cycle-weight case} of the general-cycle model is considered:  the only
restriction on $\alpha_\ell$ is that $\alpha_\ell$ goes to zero in $\ell$
faster than $1/\log\ell$.  The \emph{Ewens model}, treated in this paper (see
also \cite{bib:ewens}), is another special case of the general-cycle model: it
has $\alpha_\ell \equiv \alpha$ constant in $\ell$.

\begin{figure}
\begin{center}
\psfragscanon
\includegraphics[scale=1.0]{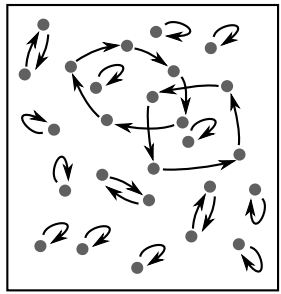}
\caption{A spatial permutation.
	\label{fig:pmt1}}
\end{center}
\end{figure}

One may hold point positions fixed, e.g. on the fully occupied unit lattice;
this approach has been taken for all simulations done up to the present by the
author and by Gandolfo \cite{bib:gru}, including specifically the work
described in this paper.  One obtains a Gibbs probability distribution on
$\SN$:
\begin{align}
\label{eqn:results_RCM_P}
	Y(\Lambda,\vecX) &= {\sum_{\sigma\in\SN} e^{-H(\vecX,\sigma)}}, &
	P(\pi) &= \frac{e^{-H(\vecX,\pi)}} {Y(\Lambda,\vecX)}.
\end{align}
(Alternatively, one may integrate over all positions in $\Lambda$, with a
resulting Gibbs distribution on $\SN$.  Here, several analytical results are
available \cite{bib:bu07,bib:bu08}.) For a random variable $S(\pi)$, we have
\begin{align}
	\bbE[S] = \sum_{\pi\in\SN} P(\pi) S(\pi).
\end{align}

\subsection{Qualitative characterization of long cycles}
\label{subsec:qual_long_cycles}

One next inquires which permutations are typical in this temperature-dependent
probability distribution on $\SN$.  In this section we develop intuition; in
the next section, we construct quantitative descriptions of the ideas presented
here.

\begin{figure}[!htb]
\psfragscanon
\begin{center}
\includegraphics[scale=0.6]{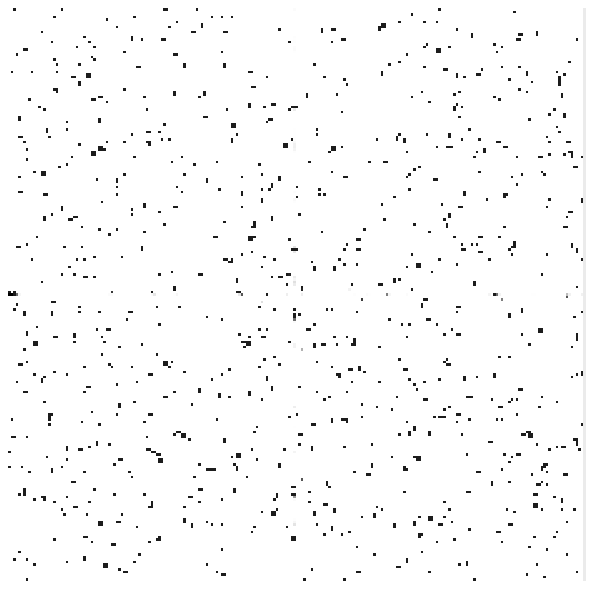}
\qquad
\includegraphics[scale=0.6]{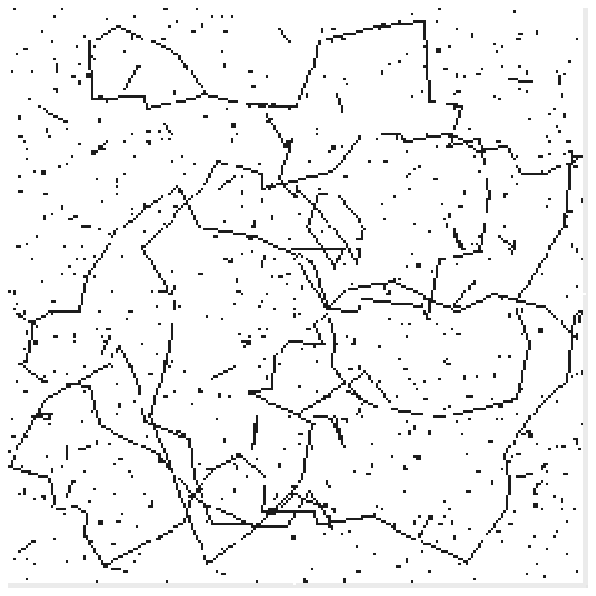}
\qquad
\includegraphics[scale=0.6]{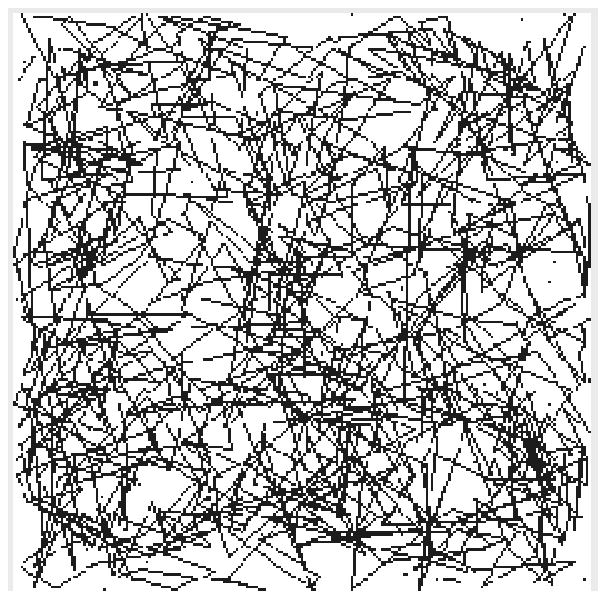}
\caption{Some typical permutations for high $T$, medium but subcritical $T$, and low $T$.
	\label{fig:dot_plots}}
\end{center}
\end{figure}

As $T \to \infty$, the probability measure becomes supported only on the
identity permutation:  the distance-dependent terms are large whenever any jump
has non-zero length.  For large but finite $T$, there are tiny islands of
2-cycles, 3-cycles, etc.  On the other hand, as $T \to 0$, length-dependent
terms go to zero, and the probability measure approaches the uniform
distribution on $\SN$:  the distance-dependent terms all go to zero.  For
intermediate $T$, one observes that the length $\|\pi(\vecx)-\vecx\|_\Lambda$
of each permutation jump remains small, increasing smoothly as $T$ drops.

\begin{figure}[!htb]
\begin{center}
\psfragscanon
\includegraphics[scale=0.6]{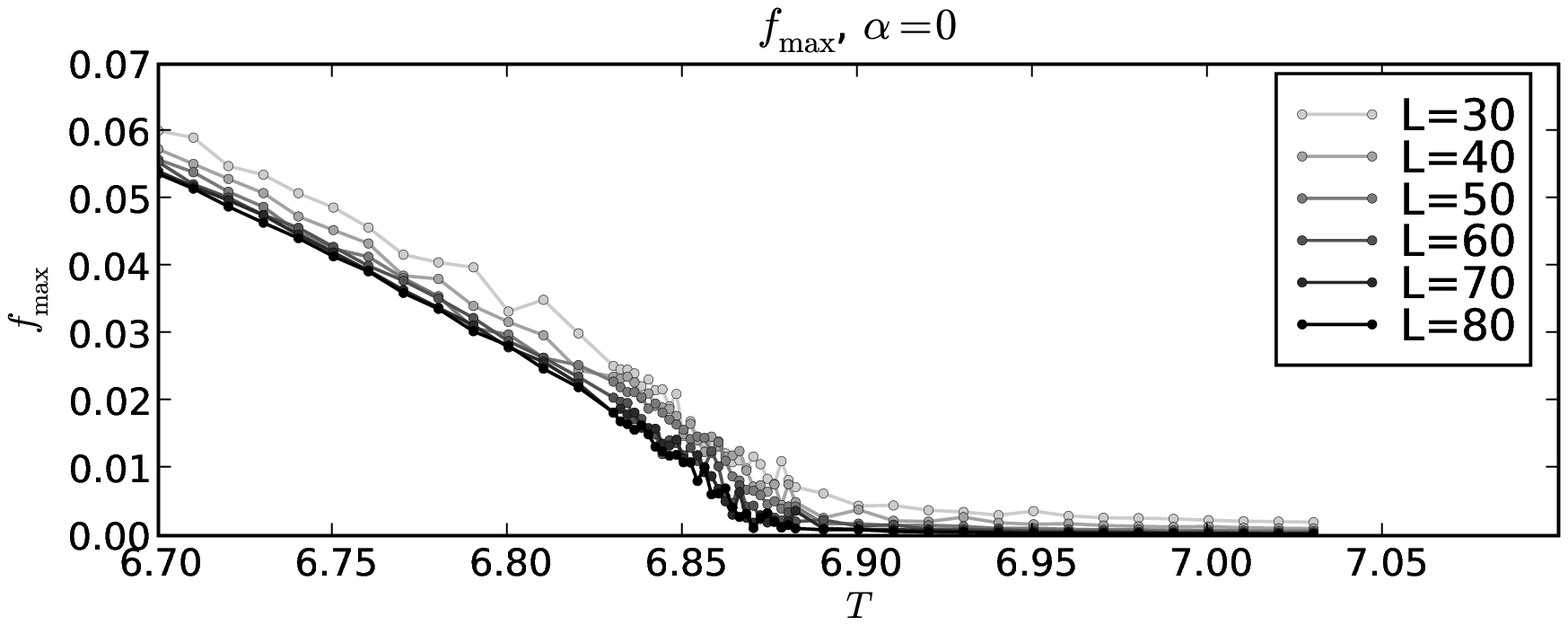}
\includegraphics[scale=0.6]{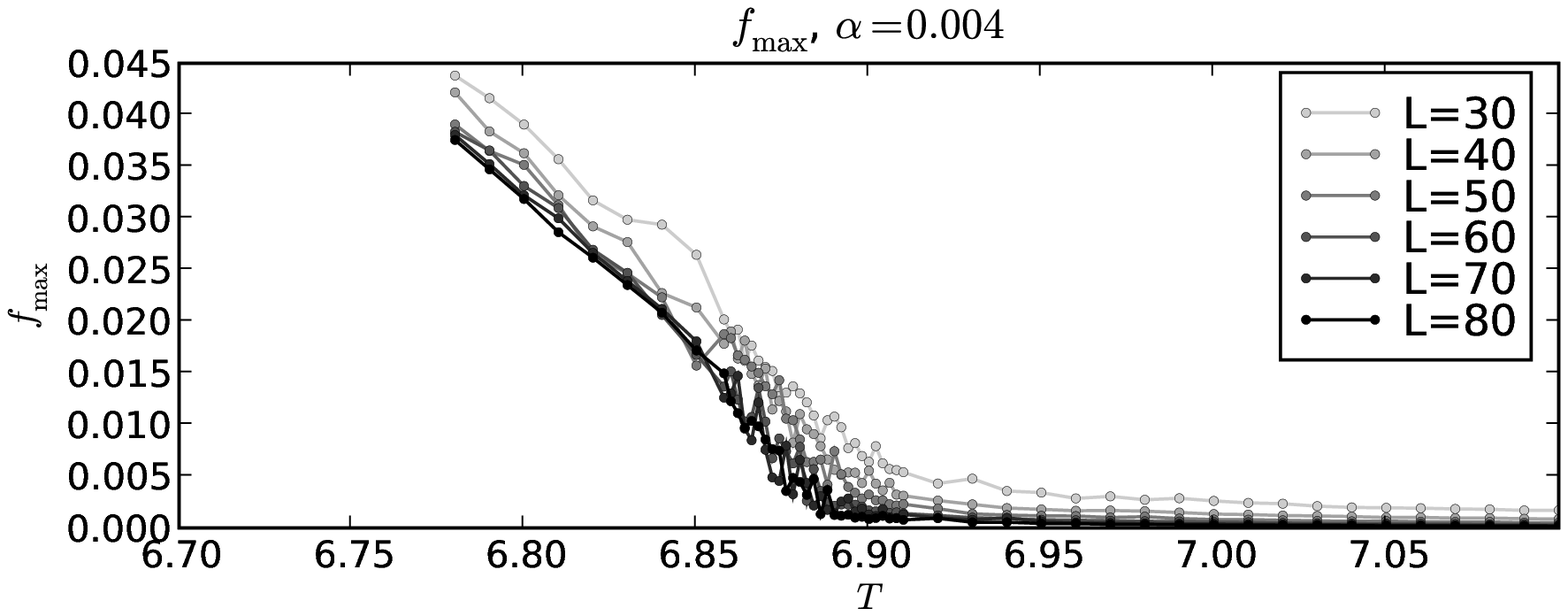}
\caption{Order parameter $f_M = \bbE[\ellmax]/N$ for finite systems, with
	$\alpha=0,0.004$.  Interactions increase the critical temperature.
	The shift is slight, but visible; we work in the regime of small
	interaction parameters.  See section \ref{sec:results} for
	a quanitative analysis of this shift.
	\label{fig:E_ellmax_N}}
\end{center}
\end{figure}

For $T$ above a critical temperature $T_c$, all cycles are short: two-cycles,
three-cycles, and so on.  We find $T_c \approx 6.86$ at $\alpha=0$, and
positive $\alpha$ terms increase $T_c$.  At $T_c$, though, there is a phase
transition: for $T<T_c$ jump lengths remain short but \emph{long cycles form}.
Quantitatively, let $\ellmax$ be the length of the longest cycle in $\pi$, with
$\bbE[\ellmax]$ its mean over all permutations.  We observe that for $T>T_c$,
$\bbE[\ellmax]$ grows only perhaps as fast as $\log(L)$ as $L\to\infty$.  That
is to say, $\bbE[\ellmax]/N$ goes to zero as $N\to\infty$.  For $T<T_c$, on the
other hand, $\bbE[\ellmax]$ scales with $N$, i.e.  $\bbE[\ellmax]/N$ approaches
a temperature-dependent constant as $N\to\infty$: there are arbitrarily long
cycles, or infinite cycles, in the infinite-volume limit.  See figure
\ref{fig:dot_plots} for depictions of typical permutations at high $T$,
subcritical $T$, and low $T$; see figure \ref{fig:E_ellmax_N} for plots of
$\bbE[\ellmax]/N$ as a function of $T$ for various system sizes with $N=L^3$.
Note in particular that higher alpha shifts the order-parameter curve to the
right, with resulting upward shift in critical temperature $T_c$.

Feynman's claim for the Bose gas is that Bose-Einstein condensation occurs if
and only if there are infinite cycles in the infinite-volume limit.  The
central point of this approach is that the system energy has been recast in
terms of permutations, which are amenable to analysis and simulation.  This
permits a new perspective on the venerable question: how does the critical
temperature of Bose-Einstein condensation depend on inter-particle interaction
strength?

Obtaining a full answer to this notoriously difficult question is a long-term
project.  As an intermediate step, we here consider the Ewens cycle-weight
Hamiltonian with point positions on the unit fully occupied unit lattice.
Through careful use of MCMC algorithms, statistical analysis, and finite-size
scaling, we are able to quantify the dependence of critical temperature on
interaction strength.

\subsection{Quantitative characterization of long cycles}
\label{subsec:ordps}

Various order parameters may be defined; all of them may be used to locate the
critical temperature $T_c(\alpha)$.
The fraction $\bbE[\ellmax]/N$ discussed above will, for brevity, be hereafter
referred to as $f_M$.
The \emph{fraction of sites in long cycles}, $f_I$, is described in detail in
\cite{bib:gru}.
The \emph{correlation length} $\xi(T)$ is defined to be the spatial length of
the cycle containing a given point $\vecx$: for $T<T_c$, it blows up as $L$
increases.  Namely, we define
\begin{align*}
	s_\vecx(\pi) &= \|\pi(\vecx)-\vecx\|_\Lambda &
	&\tand&
	s(\pi) &= \frac{1}{N} \sum_{i=1}^N s_{\vecx_i}(\pi).
\end{align*}
The expectation over all $\pi$ of $s_\vecx$ is the same as $s$, of course;
in a Monte Carlo simulation, however, the latter yields a larger sample
size and thus a smaller error bar.  We use $\xi(T) = \bbE[s]$.

\emph{Winding numbers} count the integer number of $x,y,z$ wraps around the
3-torus ($\Lambda$ with periodic boundary conditions).  Specifically, the
winding number of a permutation $\pi$ is the triple
\begin{align}
\label{eqn:def_wno}
	\vecW &= (W_x, W_y, W_z) = \frac{1}{L} \sum_{i=1}^N
		\torusdiff(\vecx_{\pi(i)}, \vecx_i),
\end{align}
where $\torusdiff$ is the \plainidx{difference vector} defined as follows.
For $\vecz \in \Lambda$, we define a \plainidx{zero-centered modulus}
vector $\vecm_L(\vecz)$.  For $\vecx,\vecy\in\Lambda$, this gives rise to a
\plainidx{difference vector} $\torusdiff(\vecx,\vecy)$:
\begin{align}
\label{eqn:vecm_L}
	\vecm_L(\vecz) &= \colvecthree{m_L(z_1)}{m_L(z_2)}{m_L(z_3)} \\
\label{eqn:n_L}
	n_L(z) &= n \in \Z \textrm{ which minimizes } |z+nL| \\
\label{eqn:m_L}
	m_L(z) &= z + n_L(z) L \\
\label{eqn:torusdiff}
	\torusdiff(\vecx,\vecy) &= \vecm_\Lambda(\vecx-\vecy).
\end{align}
We also write
\begin{align*}
	\vecW^2 &= \vecW \cdot \vecW = W_x^2 + W_y^2 + W_z^2.
\end{align*}

The \emph{scaled winding number} \cite{bib:pc87} is
$$
	f_S=\frac{\pig{\vecW^2} L^2}{3\beta N}.
$$
Lastly, the order parameter $f_W$ is the fraction of sites which participate in
winding cycles.

The order parameters $f_I(T)$, $f_S(T)$, and $f_W(T)$ show behavior similar to
$f_M := \bbE[\ellmax]/N$ (figure \ref{fig:E_ellmax_N}):  asymptotically as
$N\to\infty$, they are zero for $T \ge T_c$ and non-zero for $T < T_c$.  For
finite $N$, the curves remain analytic:  finite-size effects persist.  The
inverse correlation length $1/\xi(T)$, on the other hand, is zero for $T \le
T_c$ and non-zero for $T > T_c$.

Our goal is to quantify the dependence of $T_c$ on $\alpha$, where
\begin{align}
\label{eqn:Delta_Tc}
	\Delta T_c(\alpha) = \frac{T_c(\alpha)-T_c(0)}{T_c(0)}.
\end{align}
Known results and conjectures are formulated quantitatively in terms of
$
	\lim_{\alpha\to 0} \Delta T_c(\alpha).
$

\subsection{Known results and conjectures}
\label{subsec:conj}

Known results for point locations averaged over the continuum are
obtained largely using Fourier methods \cite{bib:bu08}, which are
unavailable for point positions held fixed on the lattice.
Betz and Ueltschi have determined $\Delta T_c(\alpha)$, to first order in
$\alpha$, for two-cycle interactions \cite{bib:bu07} and decaying cycle weights
\cite{bib:bu08}.  (This taps into a long and controversial history in the
physics literature: see \cite{bib:baymetal} or \cite{bib:su09} for surveys.)
The critical $(\rho,T,\alpha)$ manifold relates $\rho_c$ to $T_c$.
Specifically,
\begin{align}
\label{eqn:rho_c}
	\rho_c(\alpha_1, \alpha_2, \ldots)
	&= \sum_{\ell \ge 1} e^{-\alpha_\ell}
		\int_{\R^3} e^{-\ell \, 4\pi^2 \beta \|\veck\|^2}\,d\veck 
	= \frac{1}{(4\pi\beta)^{3/2}}
		\sum_{\ell \ge 1} e^{-\alpha_\ell} \ell^{-3/2} \\
	\Delta T_c(\alpha) &= c \rho^{1/3} \alpha, \quad
	\textrm{for}\; \alpha \approx 0.
\end{align}
Using this formula for constant cycle weights $\alpha_\ell \equiv \alpha$ and
for lattice density $\rho=1$, we have
\begin{equation}
\begin{aligned}
\label{eqn:theo_Tc}
	\rho_c &= \frac{\zeta(3/2) e^{-\alpha}}{(4\pi\beta)^{3/2}},
	&
	T_c &= \frac{4 \pi e^{2\alpha/3}}{\zeta(3/2)^{2/3}}
		\approx 6.626 e^{2\alpha/3}, \\
	\Delta T_c(\alpha) &= \frac{T_c(\alpha) - T_c(0)}{T_c(0)}
	= e^{2\alpha/3}-1 \approx \frac{2\alpha}{3},  &
	c &\approx 0.667.
\end{aligned}
\end{equation}
We inquire whether this result, obtained for decaying cycle weights with point
positions varying on the continuum, holds for Ewens weights with point
positions held fixed on the lattice.

For $\alpha_\ell \equiv 0$ (the non-interacting model), $\bbE[\ellmax] / N f_I$
is constant for $T$ below but near $T_c$.  (That is, the two order parameters
$f_I$ and $\bbE[\ellmax]/N$ have the same critical exponent.)  For
uniform-random permutations (Shepp and Lloyd 1966 \cite{bib:shepplloyd} solved
Golomb's 1964 question \cite{bib:golomb}), $\bbE[\ellmax]/N \approx 0.6243$;
unpublished work of Betz and Ueltschi has found $\bbE[\ellmax]/N f_I$ is that
same number for the non-interacting case $\alpha_\ell \equiv 0$.  The intuition
is that long cycles are uniformly distributed within the zero Fourier mode.
(This was proved in section 5 of \cite{bib:suto1}.  Other results on the
distribution of the length and number of cycles for probabilities depending
only on the conjugacy class can be found in sections 2 and 5 of
\cite{bib:suto1}, and in \cite{bib:suto2}.)  We conjecture that $\bbE[\ellmax]
/ N f_I$ is $\alpha$-dependent but constant in $T$ (for $T$ below but near
$T_c$) for all interaction models.

We suspect that the fine details of point positions are unimportant for the
shift in critical temperature.  Thus, $\Delta T_c(\alpha)$ on the lattice
should be similar to that on the continuum, if decaying cycle weights are used.
For Ewens interactions, though, $\Delta T_c(\alpha)$ is theoretically unknown
for Ewens interactions with points either on the continuum or on the lattice.
The simulational treatment in this paper is the only known attack on this
question.

\section{Simulational methods}
\label{sec:simulational_methods}

We run Markov chain Monte Carlo experiments for various values of $L$, $T$, and
interaction strength $\alpha$.  For each parameter combination, we generate $M$
typical permutations $\pi_1, \ldots, \pi_M$ from the stationary distribution,
using MCMC algorithms described below, and we compute random variables $X_i =
X(\pi_i)$.  (The values of $M$ used are $10^5$ away from $T_c$, and $10^6$ near
$T_c$ where sample variance is higher.) We find the sample mean and estimate
the variance of the sample mean.  The correlation of the $X_i$'s complicates
the latter.  Finite-size scaling compensates for finite-size effects:
mathematically, we are interested in estimating infinite-volume quantities
based on finite-volume numerical experiments.

\subsection{The swap-only algorithm}
\label{subsec:SO_algorithm}

Recall from section \ref{subsec:prb_model} that the expectation of a random
variable
$S$ (such as $\xi$, $f_M$, $f_W$, $f_I$, $f_S$) is
$$
	\bbE[S] = \sum_{\pi\in\SN} P(\pi) S(\pi).
$$
The number of permutations, $N!$, grows intractably in $N$.  As is typical in
Markov chain Monte Carlo methods \cite{bib:berg,bib:landaubinder}, one contents
oneself with a smaller number of samples: the expectation is instead estimated
by summing over some number $M$ ($10^5$ or $10^6$) of typical permutations.

\begin{figure}[!htb]
\begin{center}
\psfragscanon
\includegraphics[scale=0.6]{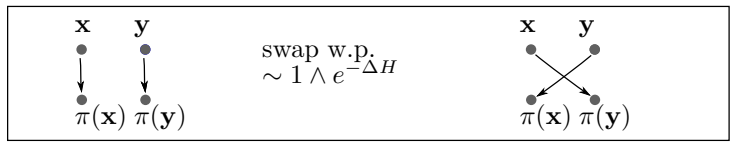}
\caption{Metropolis moves for the swap-only algorithm.
	\label{fig:gkswap}}
\end{center}
\end{figure}

The \emph{swap-only algorithm} for sampling from the Gibbs distribution
(equation \eqref{eqn:results_RCM_P}) is as follows:
\begin{itemize}
\item Start with the identity or uniform-random permutation.
\item Sweep through sites $\vecx$ of the lattice in either lexical or
uniform-random order.
\item For each site $\vecx$, do a \emph{Metropolis step}:
	\begin{itemize}
	\item Choose a site $\pi(\vecy)$ from among the six nearest
	neighbors of $\pi(\vecx)$.
	\item Propose to change $\pi$ to the permutation $\pi'$ which
		has $\pi'(\vecz)=\pi(\vecz)$ for all $\vecz \ne \vecx, \vecy$
		but $\pi'(\vecx) = \pi(\vecy)$ and $\pi'(\vecy) = \pi(\vecx)$.
		(See figure \ref{fig:gkswap}.)
	\item With probability proportional to $\min\{1, \edh\}$
	where $\Delta H = H(\pi') - H(\pi)$, accept the change.
	(If the change is rejected, $\pi'=\pi$.)
	\end{itemize}

\item After each sweep, obtain a value of each random variable for inclusion in
computation of its sample mean.

\end{itemize}

One starts accumulating data only after a suitable number of
\emph{thermalization sweeps}.  The idea is that the initial, identity
permutation is not typical, nor are the first few afterward.  The integrated
autocorrelation time \cite{bib:berg} of system energy $H$ gives an idea of how
many Metropolis sweeps should be discarded before the permutations become
typical.  Also, one may examine $H$ to ensure that it has reached its long-term
average value.  This is explained in detail in \cite{bib:kerl}.  We next prove
correctness of this algorithm.

\subsection{Explicit construction of the Markov matrix}
\label{subsec:SAR_Markov_matrix}

For section \ref{subsec:SO_correctness} we will need an explicit construction of
the Markov matrix corresponding to the swap-only algorithm as described in
section \ref{subsec:SO_algorithm}.  The Markov perspective on the algorithm is
that the distribution $P^{(0)}(\pi)$ of the first permutation is either
supported solely on the identity, or uniform on all $N!$ permutations.  The
distribution for subsequent permutations is
\begin{align*}
	P^{(k+1)}(\pi') &= \sum_{\pi \in \SN} P^{(k)}(\pi) M(\pi, \pi')
\end{align*}
or, in matrix/vector notation,
\begin{align*}
	\vecP^{(k+1)} &= \vecP^{(k)} \vecM.
\end{align*}
In this section we precisely describe the matrix $\vecM$; in section
\ref{subsec:SO_correctness} we show that $\vecP^{(k)}$ approaches the
Gibbs distribution (equation \eqref{eqn:results_RCM_P}).

The matrix $\vecM$ is $N! \times N!$: rows are indexed by $\pi_1, \ldots,
\pi_{N!}$ and columns are indexed by $\pi'_1, \ldots, \pi'_{N!}$.  Most of the
entries of $\vecM$ are zero:  Metropolis steps change only two permutation
sites whereas most $\pi,\pi'$ differ at more than two sites.

\begin{defn}
For $\pi,\pi'\in\SN$, define
\begin{align*}
	d(\pi,\pi') &= \#\{i=1,2,\ldots,N : \pi(i) \ne \pi'(i)\}.
\end{align*}
\end{defn}

\begin{rem*}
Note that $d(\pi,\pi') \ne 1$ since if two permutations agree on $N-1$ sites,
they must agree on the remaining site.  It is easily shown that the function
$d(\pi,\pi')$ is a \plainidx{metric} on $\SN$.
\end{rem*}

\begin{defn}
Lattice sites $\vecx,\vecy$ are \emph{nearest-neighbor} if
$\|\vecx-\vecy\|_\Lambda = 1$.
\end{defn}

\begin{defn}
\label{defn:R_pi}
For $\pi\in\SN$, define
\begin{align*}
	R(\pi) &= \{\pi'\in\SN: d(\pi,\pi')=2 \textrm{ and }
		\|\pi(\vecx)-\pi(\vecy)\|_\Lambda=1\}
\end{align*}
where the $\vecx$ and $\vecy$ are taken to be the two points at which
$\pi,\pi'$ differ.  Then $R(\pi)$ is the set of permutations $\pi'$ reachable
from $\pi$ on a swap.
\end{defn}

We construct the Markov matrix for use when sites $\vecx$ are selected at
uniform random.  (The matrices for use when $\vecx$ is selected sequentially
are similar.) For each $\pi\in\SN$,
\begin{align}
\label{eqn:SO_M}
	M(\pi,\pi') &= \begin{cases}
		\frac{1}{3N}\left(1 \wedge e^{-H(\pi')+H(\pi)}\right),
		& \pi' \in R(\pi), \\
		\displaystyle
		1 - \sum_{\pi'' \in R(\pi)}
		\frac{1}{3N}\left(1 \wedge e^{-H(\pi'')+H(\pi)}\right),
		& \pi=\pi'; \\
		0,
		& \textrm{ otherwise}.
	\end{cases}
\end{align}
To justify the choice of prefactor $1/3N$, note that there are $N$ choices of
lattice points $\vecx$.  For each $\vecx$, there are 6 choices of $\piy$ which
are nearest neighbors to $\pix$.  This double-counts the $3N$ distinct choices
of $\pi'$ reachable from $\pi$ in a single Metropolis step, since choosing
$\vecx$ and then $\vecy$ results in the same Metropolis step as choosing
$\vecy$ and then $\vecx$.

\subsection{Correctness of the swap-only algorithm}
\label{subsec:SO_correctness}

It is clear that the swap-only algorithm produces a sequence of permutations,
but with what distribution?  From Markov-chain theory, we know the following:
If the chain is irreducible, aperiodic, and satisfies detailed balance, then
the chain has the Gibbs distribution (equation \eqref{eqn:results_RCM_P}) as its
unique invariant distribution.

We note the following terminology: \emph{detailed balance} is the same as
\emph{reversibility}\index{reversible}.  Also, an irreducible, aperiodic chain
on a finite state space is called \emph{ergodic}.  Also note from Markov-chain
theory that all states in a recurrence class have the same period.  Thus, if we
can show that the chain is irreducible (i.e. the entire state space is a single
recurrence class), then for aperiodicity of the chain it suffices to show that
a single state (e.g. the identity permutation) has period 1.

\begin{prop}[Irreducibility]
\label{prop:SO_irreducible}
For all $\pi,\pi'$, there is an $n$ such that $M^n(\pi,\pi') > 0$.
That is, any permutation is reachable from any other.
\begin{proof}
Transpositions generate $\SN$:  for all $\pi\in\SN$, there exist transpositions
$\sigma_1, \ldots, \sigma_m$ such that
$	\pi = \prod_{j=1}^m \sigma_j$.
Thus, it suffices to show that given any permutation $\pi$ and any two points
$\vecx$ and $\vecz$, so $\pi: \vecx \mapsto \pi(\vecx)$ and $\pi: \vecz \mapsto
\pi(\vecz)$, we can construct a sequence of swaps sending $\pi$ to $\pi'$ so
that $\pi': \vecx \mapsto \pi(\vecz)$, $\pi': \vecz \mapsto \pi(\vecx)$, and
$\pi'(\vecy) = \pi(\vecy)$ for all $\vecy \ne \vecx,\vecz$.  (If $\pi(\vecx)$
and $\pi(\vecz)$ are nearest-neighbor lattice sites, of course, then a single
swap does the job.)

Define $G_{\veca,\vecb}:\SN\to\SN$ to be the swap operator for nearest-neighbor
lattice sites $\pi(\veca)$ and $\pi(\vecb)$.  Write $\pi'=G_{\veca,\vecb} \pi$.
Given $\vecx$ and $\vecz$, there is a (non-unique) sequence of lattice sites
$\vecy_0, \vecy_1, \vecy_2, \ldots, \vecy_n$ such that $\vecy_0=\vecx$,
$\vecy_n=\vecz$, and $\|\pi(\vecy_{i+1})-\pi(\vecy_i)\|_\Lambda=1$ for $i=0, 1,
\ldots, n-1$.  (See figure \ref{fig:irrprop}.)  We will construct a sequence of
swaps along this nearest-neighbor path whose end result is to swap the
permutation arrows starting at $\vecx$ and $\vecz$, leaving all other arrows
unchanged.  We first need a lemma about compositions of swaps.

\begin{figure}[!htb]
\begin{center}
\psfragscanon
\includegraphics[scale=0.6]{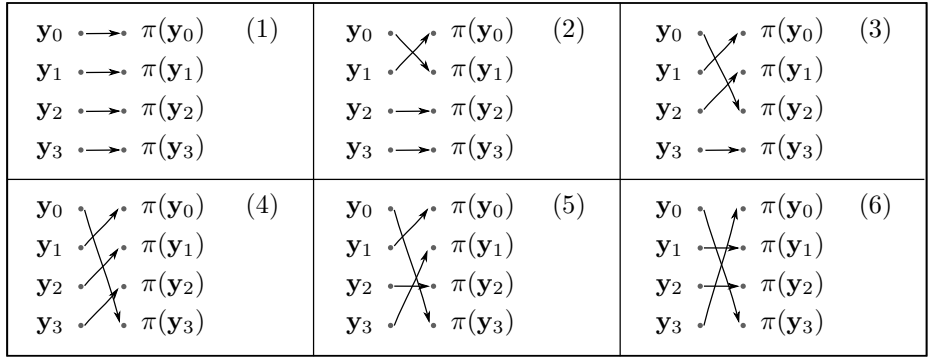}
\caption{A sequence of (nearest-neighbor) swaps which results in a
	non-nearest-neighbor swap.
	\label{fig:irrprop}}
\end{center}
\end{figure}

\begin{notn}
Given $\vecx_1, \ldots, \vecx_N$ and a permutation $\pi$, we may write
$\pi$ as an \emph{image map} with the $\vecx_i$'s along the top row and
their images along the bottom row:
\begin{align*}
	\pcmatrix{\vecx_1 & \ldots & \vecx_N \\
	\pi(\vecx_1) & \ldots & \pi(\vecx_N)}
\end{align*}
\end{notn}

We find that the composition of maps
\begin{align}
	(
	G_{\vecy_n,\vecy_1} \circ
	G_{\vecy_n,\vecy_2} \circ \ldots \circ
	G_{\vecy_n,\vecy_{n-2}} \circ
	G_{\vecy_n,\vecy_{n-1}}
	)\circ(
	G_{\vecy_0,\vecy_n} \circ
	G_{\vecy_0,\vecy_{n-1}} \circ \ldots \circ
	G_{\vecy_0,\vecy_2} \circ
	G_{\vecy_0,\vecy_1}
	)
\end{align}
swaps the images of $\vecx=\vecy_0$ and $\vecz=\vecy_n$ while leaving
all other images unchanged, that is,
\begin{align*}
	\pcmatrix{
	    \vecy_0  &     \vecy_1  & \ldots&     \vecy_{n-1}  &     \vecy_n \\
	\pi(\vecy_0) & \pi(\vecy_1) & \ldots& \pi(\vecy_{n-1}) & \pi(\vecy_n)
	}
	\mapsto
	\pcmatrix{
	    \vecy_0  &     \vecy_1  & \ldots&     \vecy_{n-1}  &     \vecy_n \\
	\pi(\vecy_n) & \pi(\vecy_1) & \ldots& \pi(\vecy_{n-1}) & \pi(\vecy_0)
	}.
\end{align*}
\end{proof}
\end{prop}

\begin{rem*}
Below we will discuss winding numbers, and the empirical fact that the
swap-only algorithm changes them only rarely.  The chain is irreducible but
various non-zero transition probabilities can still be very small.
\end{rem*}

\begin{defn}
\label{defn:period}
The \emphidx{period} of $\pi$ is
\begin{align*}
	p(\pi) &= \gcd\{n: P(\Pi_n=\pi \mid \Pi_0=\omega) > 0\}
\end{align*}
where $\Pi_k$ is the random variable which is the permutation appearing at the
$k$th step of the Markov chain.  We say that $\pi$ has period $p$ if it
reappears with probability 1 after every $p$ steps.  A permutation $\pi$ is
\emphidx{aperiodic} if $p(\pi)=1$.  The chain is \emphidx{aperiodic} if
$p(\pi)=1$ for every $\pi$.
\end{defn}

\begin{prop}[Aperiodicity]
\label{prop:SO_aperiodic}
The swap-only algorithm's Markov chain is \plainidx{aperiodic}.
\begin{proof}
This follows from irreducibility, which says in particular that for every
$\pi$, there is an integer $m$ such that $M^m(\pi,\pi) > 0$.  Then
$M^{n}(\pi,\pi)>0$ for all $n>m$, implying $p(\pi)=1$.
\end{proof}
\end{prop}

\begin{prop}[Detailed balance]
\label{prop:SO_DB}
For all $\pi,\pi' \in \SN$,
\begin{align}
\label{eqn:SO_DB}
	P(\pi) M(\pi, \pi') &= P(\pi') M(\pi', \pi).
\end{align}
\begin{proof}
The detailed-balance statement in terms of the Gibbs distribution (equation
\eqref{eqn:results_RCM_P}) and the Metropolis transition matrix (equation
\eqref{eqn:SO_M}) is
\begin{align*}
	\frac{e^{-H(\pi)}}{Z}
	\left(1 \wedge e^{-H(\pi')} e^{H(\pi)}\right)
	&\maybeeq
	\frac{e^{-H(\pi')}}{Z}
	\left(1 \wedge e^{-H(\pi)} e^{H(\pi')}\right).
\end{align*}
The $Z$'s cancel.
The lemma below shows that $M(\pi,\pi') \ne 0$ iff $M(\pi',\pi) \ne 0$.
If $M(\pi,\pi') = 0$, then detailed balance holds.
If $M(\pi,\pi') \ne 0$, then there are two cases.
If $H(\pi') \le H(\pi)$, then
\begin{align*}
 	e^{-H(\pi)} \left(1\right)
	&=
	e^{-H(\pi')} \left(e^{-H(\pi)} e^{H(\pi')}\right).
\end{align*}
If $H(\pi') > H(\pi)$,
\begin{align*}
	e^{-H(\pi)} \left(e^{-H(\pi')} e^{H(\pi)}\right)
	&=
 	e^{-H(\pi')} \left(1\right).
\end{align*}
In all cases, detailed balance holds.
\end{proof}
\end{prop}

\begin{lem}
\label{lem:nzdb1}
For all $\pi,\pi' \in \SN$,
\begin{align*}
	M(\pi,\pi') \ne 0 \iff M(\pi', \pi) \ne 0.
\end{align*}
\begin{proof}
This is true since $\pi' \in R(\pi)$ if and only if $\pi \in R(\pi')$, which is
a direct consequence of the definition \ref{defn:R_pi} of $R(\pi)$.
\end{proof}
\end{lem}

This lemma completes the proof that the swap-only algorithm satisfies detailed
balance and thus has the Gibbs distribution as its invariant distribution.  It
is not hard to show that if swaps sites $\vecx\ne \vecy$ are in the same cycle
before a swap, they are in different cycles after the swap, and vice versa.
This is not a correctness result, but rather a sanity check:  it shows that
cycles may grow or shrink upon swap-only moves.

\begin{figure}[!htb]
\begin{center}
\psfragscanon
\includegraphics[scale=0.7]{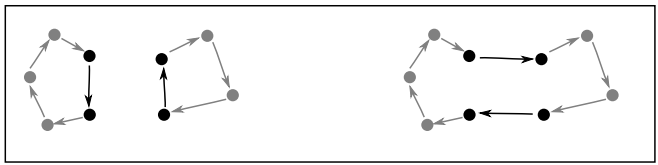}
\caption[Swaps merge disjoint cycles and split single cycles.]
	{Swaps merge disjoint cycles and split single cycles.
	The left-hand permutation can be reached from the right-hand permutation
	via a swap, and vice versa.
	\label{fig:gkcycles}}
\end{center}
\end{figure}

\subsection{Winding cycles and the swap-and-reverse algorithm}
\label{subsec:wno_SAR}

The propositions of section \ref{subsec:SO_correctness} showed that the
swap-only algorithm is correct --- in particular, any permutation is reachable
from any other with non-zero probability.  However, in practice some of these
non-zero transition probabilities can be quite small.  In particular, we
observe that the swap-only algorithm almost always generates permutations with
zero winding number.

\begin{figure}
\begin{center}
\psfragscanon
\includegraphics[scale=0.9]{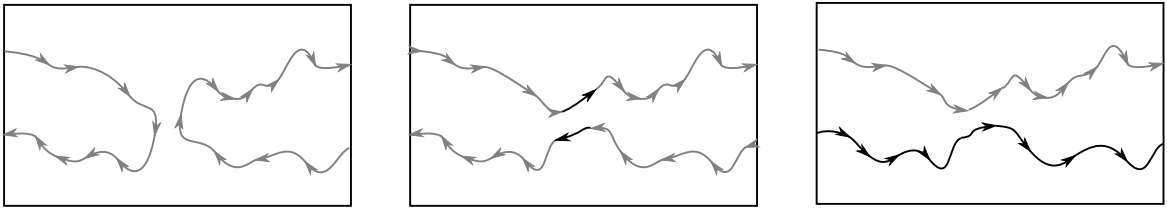}
\caption{Conservation of winding number in the swap-only algorithm.
	\label{fig:gkwinding}}
\end{center}
\end{figure}

This problem, and a partial solution, is explained intuitively by figure
\ref{fig:gkwinding} and rigorously in \cite{bib:kerl}.  Part 1 of the figure
shows a permutation $\pi$ with a long cycle on the torus which almost meets
itself in the $x$ direction.  In part 2, after a Metropolis step sending $\pi$
to $\pi'$, one cycle winds by $+1$ and the other by $-1$.  Metropolis steps
create winding cycles only in opposite-direction pairs; the total $W_x(\pi)$ is
still zero.  Part 3 of the figure shows that if we reverse one cycle (which is
a zero-energy move), $W_x(\pi)$ is now 2.  In general (with full details in
\cite{bib:kerl}), winding numbers of even parity can be generated.

Our current best algorithm (swap-and-reverse) has two types of sweeps:  (1) For
each lattice site, do a Metropolis step as above.  (2) For each cycle in the
permutation, reverse the direction of the cycle with probability $1/2$.  This
permits winding numbers of even parity in each of the three axes.

We have experimented with various methods to obtain winding numbers of all
parities.  The creation or destruction of a winding cycle is a non-local
update; one is reminded of the Swendsen-Wang algorithm for the Ising model.
However, our attempt at non-local updates has an unreasonably low acceptance
rate, namely, on the order of $e^{-L}$ where $L$ is the box length.

We have also created a \emph{worm algorithm}, inspired by approaches to this
same winding-number problem in path-integral Monte Carlo methods
\cite{bib:bps06,bib:pst98}.  That is, a permutation loop is selected at random
and then cut open at a randomly selected point.  The resulting worm is allowed
to move around $\Lambda$ via Metropolis moves; eventually, it closes again.
This worm algorithm has an elegant theory and correctness proof
\cite{bib:kerl}; yet, it has an unacceptably long stopping time for loop
closure, and none of our attempts to remedy the stopping-time problem have
satisfied detailed balance.

At present, we content ourselves with the swap-and-reverse algorithm; it is
used to generate all the results discussed in section \ref{sec:results}.  The
order parameters $f_S$ and $f_W$ depend on winding phenomena, but the other
three, $1/\xi$, $f_I$, and $\fM$, do not; furthermore, results obtained in
section \ref{sec:results} using each of the five order parameters are, for the
most part, compatible.  Yet, as we will see, $f_S$ and $f_W$ do not permit
successful finite-size scaling.

\subsection{Finite-size scaling}
\label{subsec:fss}

Finite-size scaling takes the form of a hypothesis, or rather a set of
hypotheses, which is tested against the data.  See also \cite{bib:cggp} for a
nice survey.

We have an infinite-volume random variable $S(T)$, e.g. any of the order
parameters defined in section \ref{subsec:ordps}.  The finite-volume quantity
is $S_L(T)$.  Define $t=(T-T_c)/T_c$.  Examine, say, $0.99 < t < 1.01$.
The first hypothesis is that the correlation length $\xi(T)$ follows a power
law
$$
	\xi(T) \sim |t|^{-\nu}, \quad T \to T_c
$$
For the infinite-volume quantity, we expect a power-law behavior
$$
	S(T) \sim t^{\rho}, (-t)^{\rho}, \qor |t|^\rho.
$$
(The domain of validity is $t<0$ or $t>0$ depending on whether the order
parameter $S$ is left-sided or right-sided, respectively.) One moreover
hypothesizes that for $T$ near $T_c$, $S_L(T)$ and $S(T)$ are related by a
\emph{universal function} $Q_S$ which depends on $T$ only through the ratio
$L/\xi$:
\begin{align}
\label{eqn:fss}
	S_L(T) &= L^{-\rho/\nu} Q_S(L^{1/\nu} t)
	\sim L^{-\rho/\nu} Q_S((L/\xi)^{1/\nu}).
\end{align}

\section{Results}
\label{sec:results}

Here we complete the steps sketched in section \ref{subsec:fss}.  The flow of
data and uncertainties are as follows:

\begin{itemize}

\item Markov chain Monte Carlo simulations, with error bars determined using
the method of integrated autocorrelation time \cite{bib:berg}, yield $S_L(T,
\alpha)$ data points.  There are five order parameters $S$, six values of $L$
(30, 40, 50, 60, 70, 80), nine values of $\alpha$, and a few dozen
values of $T$ for each $\alpha$.

\item CPU time per $L,T,\alpha$ experiment, with $10^5$ Metropolis sweeps, is
approximately 1.3 hours for $L=40$ and 20 hours for $L=80$.  For the work
described in this paper and in \cite{bib:kerl}, a total of 5.4 CPU years was
used.

\item For each $S$, $L$, and $\alpha$, we use $S_L(T,\alpha)$ values for all
available values of $T$ and $\alpha$ to estimate $\rhohat_S(L)$.  (Critical
exponents are assumed to be independent of $\alpha$ for small $\alpha$, or with
weak enough dependence on $\alpha$ that that dependence is lost in the noise.)
Error bars may be propagated from the MCMC simulations, or computed from
regression uncertainties.

\item Extrapolating $\rhohat_S(L)$ in $L \to \infty$ results in the five
estimated critical exponents $\rhohat_S$.  Uncertainties are computed from the
regression analysis.

\item Once the critical exponents are estimated, we obtain
$\hat{T}_{c,S}(\alpha)$ for each of the five order parameters $S$ and for each
$\alpha$.  Uncertainties are computed by visual inspection of the crossing
plots discussed in section \ref{subsec:crossing}.

\item Once the critical exponents and $T_c$ are known, one should be able to
obtain plots of the universal function $Q_S$ which is, up to sampling
variability, independent of $L$, $T$, and $\alpha$.  This verifies the
finite-size-scaling hypothesis.

\item The shift in reduced critical temperature is as in equation
\eqref{eqn:Delta_Tc}.  Error bars are computed from regression uncertainties.

\end{itemize}

\subsection{Determination of $L$-dependent critical exponents}
\label{subsec:rho_hat_L}

\begin{figure}
\begin{center}
\psfragscanon
\includegraphics[scale=0.50]{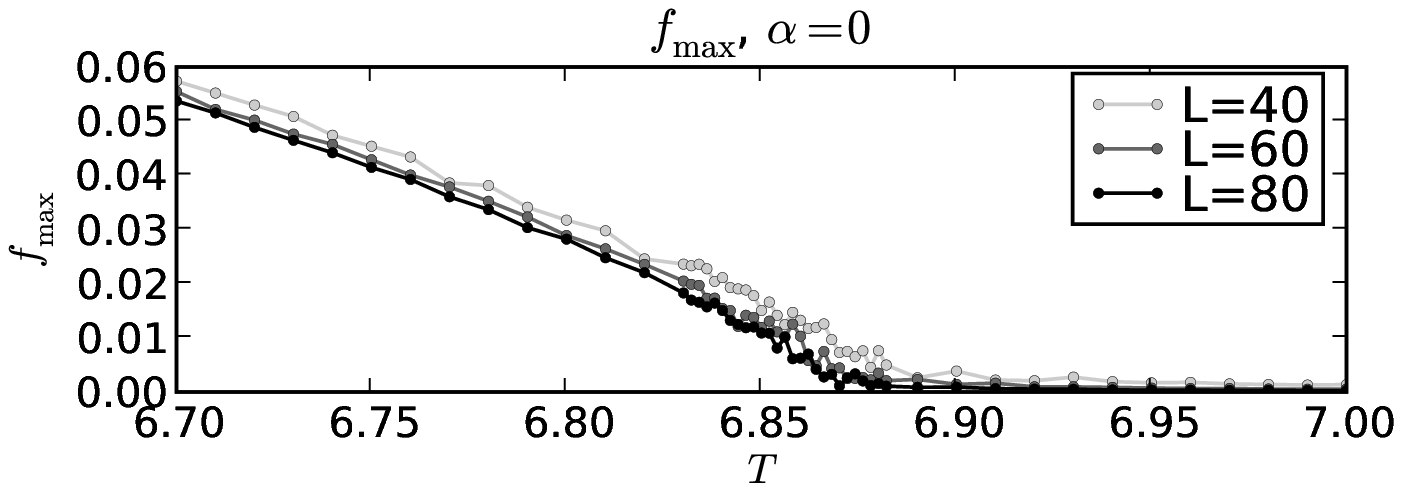}
\includegraphics[scale=0.50]{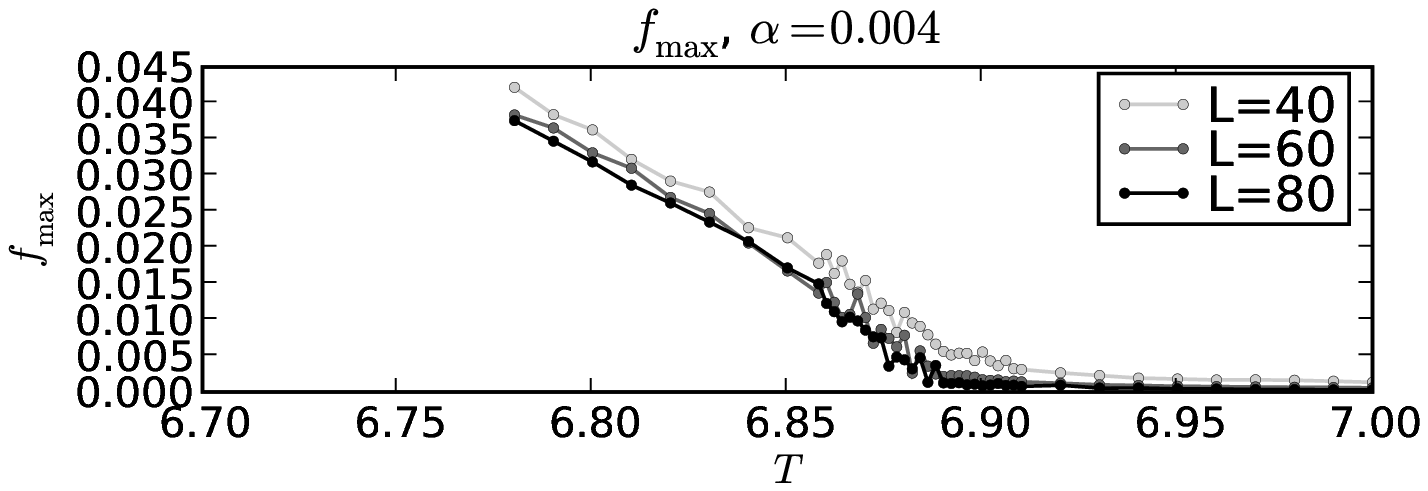}
\includegraphics[scale=0.50]{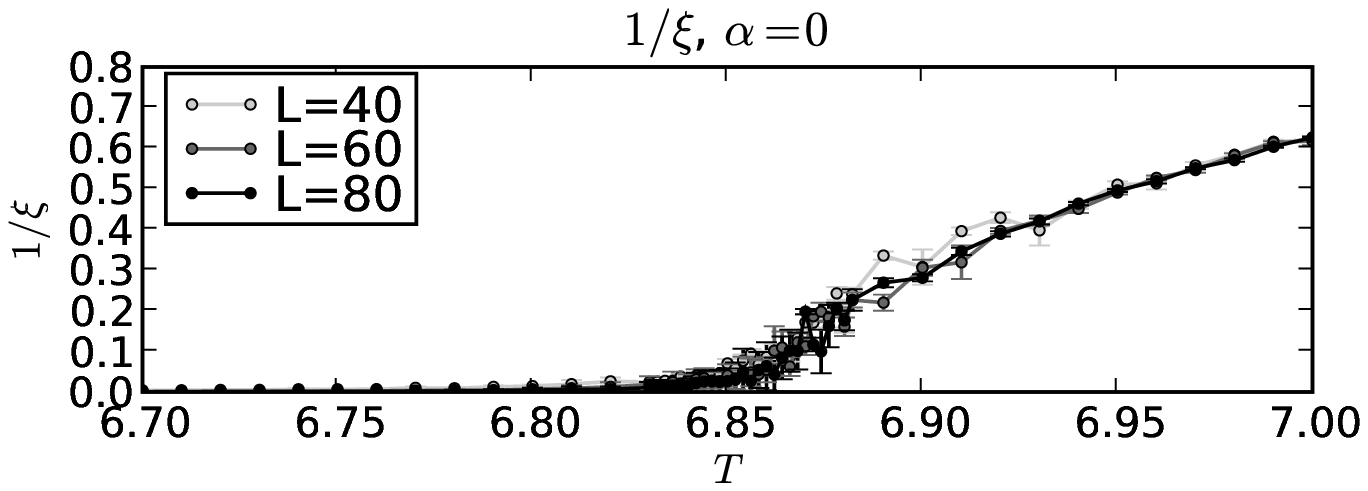}
\includegraphics[scale=0.50]{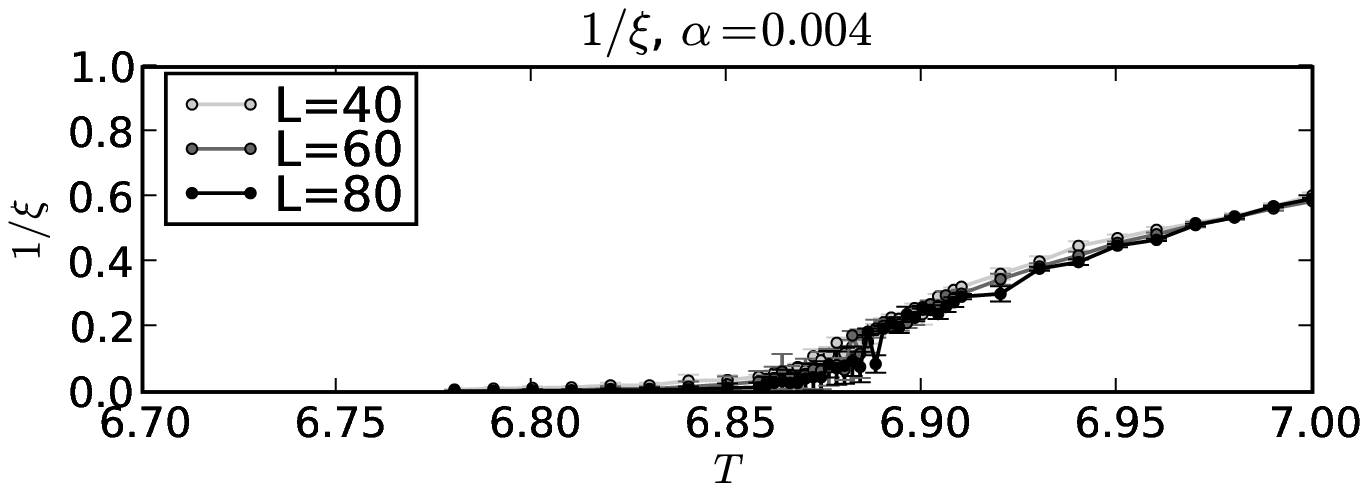}
\caption{Order parameters $f_M$ and $1/\xi$ for $L=40,60,80$ and $\alpha=0$
and $0.004$.  The remaining order parameters $f_S$, $f_W$, and $f_I$ behave
similarly to $f_M$ but with not all with the same critical exponents.
	\label{fig:f_M_and_recip_xi}}
\end{center}
\end{figure}

\begin{figure}
\begin{center}
\psfragscanon
\includegraphics[scale=0.50]{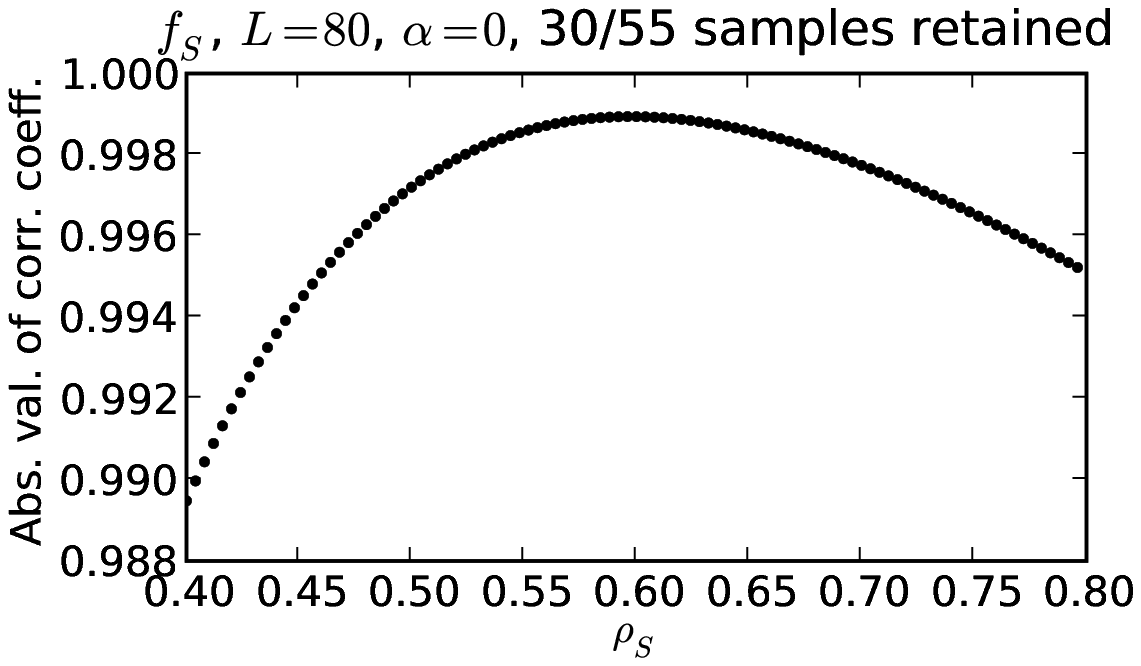}
\includegraphics[scale=0.50]{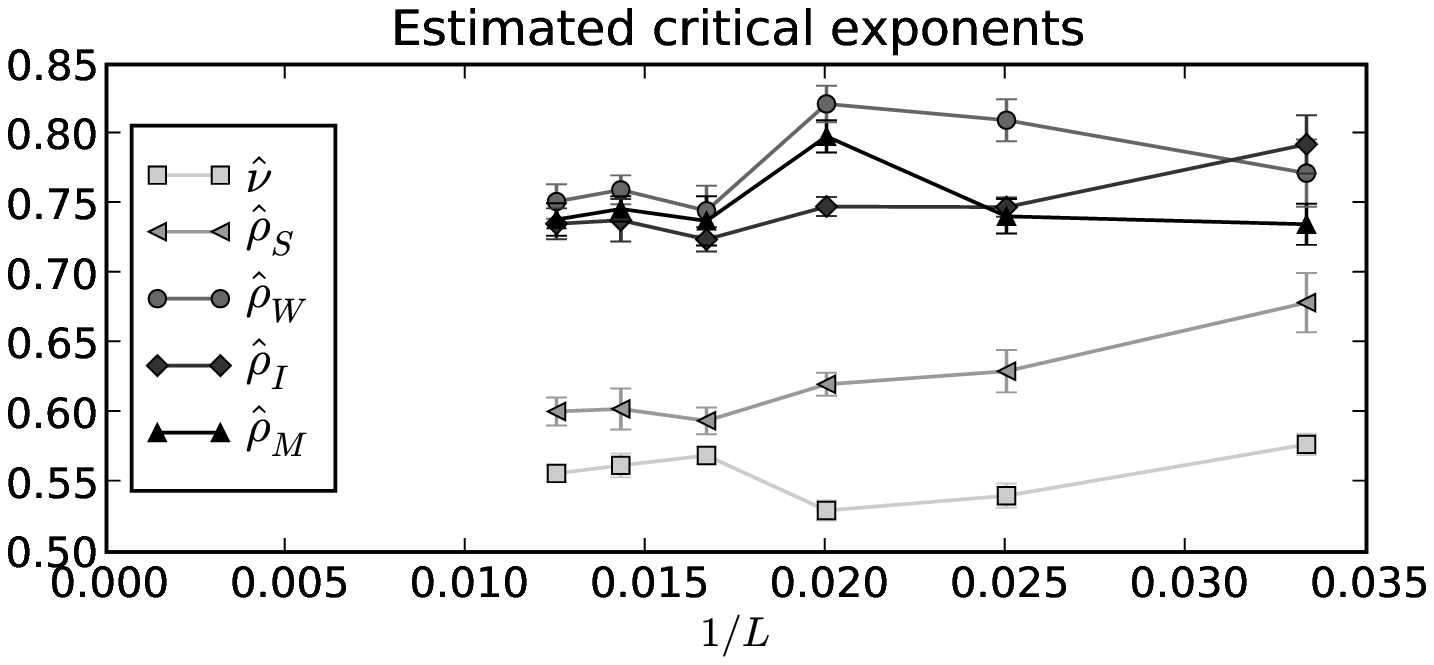}
\caption[Determination of critical exponent.]
	{On the left:  determination of critical exponent
	$\hat{\rho}_S(L,\alpha)$ for order parameter $f_S$, as the value which
	minimizes linear-regression error for $S_L(T,\alpha)^{1/\rho}$.  Visually,
	one sees $\hat{\rho}_S(L=80, \alpha=0.0) \approx 0.59$.
	On the right:  estimated critical exponents for $L=30,40,50,60,70,80$.
	\label{fig:smiley_and_rhoestalpha}}
\end{center}
\end{figure}

\begin{figure}
\begin{center}
\psfragscanon
\includegraphics[scale=0.5]{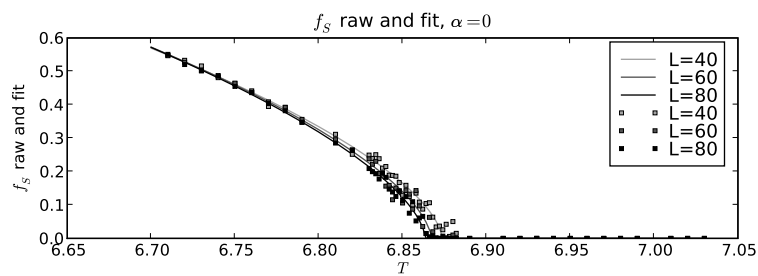}
\caption[Power-law fit vs. simulational order-parameter data.]
	{Power-law fit vs. simulational data for order parameter $f_S$,
	$\alpha=0$.
	\label{fig:ordp_fit_comparison}}
\end{center}
\end{figure}

For each of order parameter $S$, interaction parameter $\alpha$, and box length
$L$, we examine all $S(L,T,\alpha)$ data for which $S > \veps$, with $\veps$
taken from the plots to ensure that we examine the portions of the curves
corresponding to non-zero order parameter in the infinite limit (see figure
\ref{fig:f_M_and_recip_xi}).  For $1/\xi$, this means $T>T_c$; for the other
four order parameters, this means $T<T_c$.  From plots such as those in figure
\ref{fig:f_M_and_recip_xi}, we choose $\veps$ to be
$0.1$  for $1/\xi$,
$0.01$ for $f_M$,
$0.01$ for $f_I$,
$0.05$ for $f_S$, and
$0.01$ for $f_W$.
For each $S$, $\alpha$, and $L$, we then apply linear regression to
$S(L,T)^{1/\rho_S}$ for varying $\rho_S$.  We find $\rhohat_S(L)$ which
optimizes the correlation coefficient \cite{bib:young} of the linear
regression.  Results are shown in figure \ref{fig:smiley_and_rhoestalpha}.
Given $\rhohat_S(L)$ along with its corresponding linear-regression parameters
$m$ and $b$, we may plot a power-law fit to the simulational data.  One such
comparison plot is shown in figure \ref{fig:ordp_fit_comparison}.


\begin{table}
\input{fMfIvalues.tex}
\caption{$f_M/f_I$ as a function of $\alpha$.  An upward trend is
	visible, but it is not pronounced.
	\label{table:fMfI_alpha}}
\end{table}

\subsection{Extrapolation of critical exponents for the infinite-volume limit}
\label{subsec:rhohat_extrap}

Next, for each $S$, given estimates $\rhohat_S(L)$ for increasing values of
$L$, we plot $\rhohat_S(L)$ versus $1/L$.  The vertical intercept of this plot
estimates the infinite-volume exponent $\rhohat_S(\alpha)$.  (See figure
\ref{fig:smiley_and_rhoestalpha}.) Results are shown in table
\ref{table:rhohat_S}.

\begin{table}
\input{rhoest.tex}
\caption{Extrapolated estimates of the infinite-volume critical exponents,
	found from the vertical intercept of figure
	\ref{fig:smiley_and_rhoestalpha}.  \label{table:rhohat_S}}
\end{table}

\subsection{Determination of critical temperature}
\label{subsec:crossing}

\begin{figure}
\begin{center}
\psfragscanon
\includegraphics[scale=0.48]{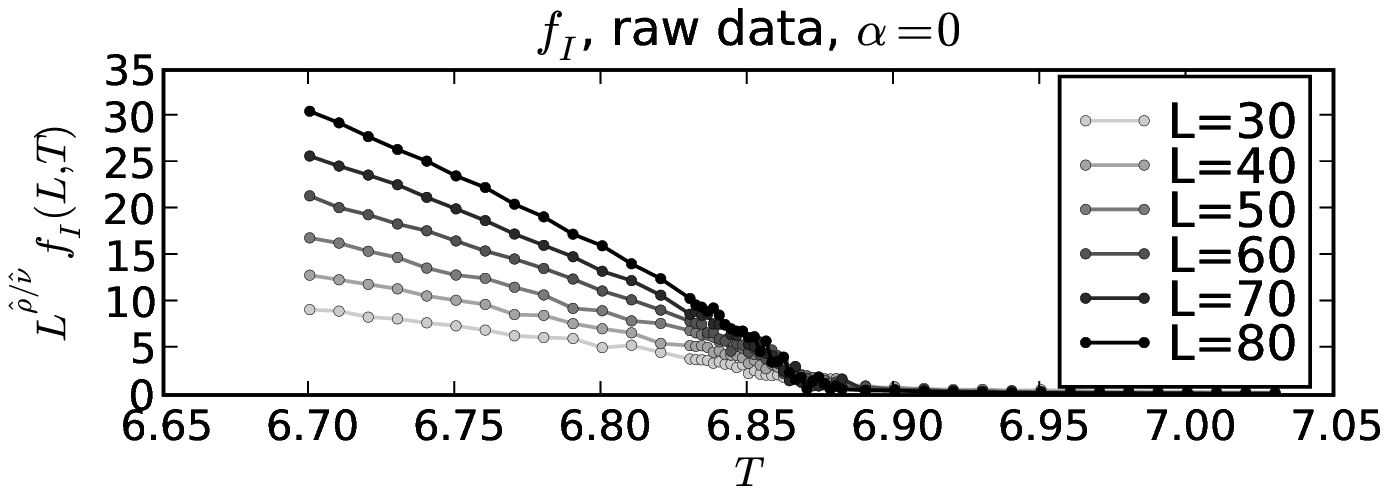}
\includegraphics[scale=0.48]{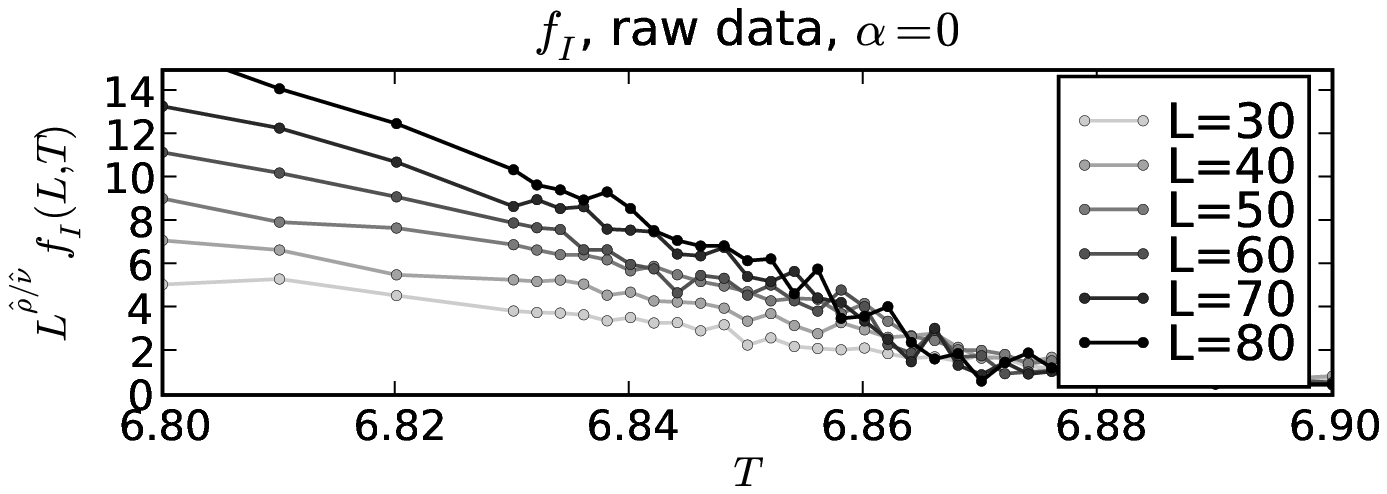}
\caption
	[Estimation of $T_c(\alpha)$ for order parameter $f_I$.]
	{The crossing method to estimate $T_c(\alpha)$ for order parameter $f_I$,
	with $\rhohat$ and $\nuhat$ as above: $T_c(\alpha)$ corresponds to the
	horizontal coordinate of the intersection point of the plots.
	The upper-right-hand plot is a close-up of the upper-left-hand plot.
	Order parameters $f_S$ and $f_W$, which depend on winding phenomena, do not
	exhibit clear crossing behavior.
	\label{fig:fS_crossing}}
\end{center}
\end{figure}

Given the above estimators of the critical exponents, the \emphidx{crossing
method} \cite{bib:cggp} estimates $T_c(\alpha)$.  Namely, we plot
$L^{\rhohat/\nuhat} S_L(T)$ as a function of $T$.  At $T=T_c$ we have $t=0$ and
$L^{\rho/\nu} S_L(T)  = Q_S(0)$, regardless of $L$ (equation \eqref{eqn:fss}).
Thus, these curves will cross (approximately, due to sampling variability) at
$T=T_c$.  If they do not, the finite-size-scaling hypothesis is not verified.
(Note in particular that for order parameter $1/\xi$ whose critical exponent is
$\nu$, we apply the crossing method to $L S_L(T)$ as a function of $T$:  thus,
the $T_c(\alpha)$ estimate using $1/\xi$ is independent of $\nuhat$.) See for
example figure \ref{fig:fS_crossing}.  (We acknowledge that larger values of
$L$, would improve the visual effect.  Results presented here are those
obtained within the timeframe of the author's doctoral dissertation work.)
Results are in figure \ref{fig:Tc_alpha}.


Using order parameters $f_S$ and $f_W$, which depend on winding phenomena, one
does not see clear crossing behavior.  We suggest that either this is related
to the even-winding-number issue discussed in section \ref{subsec:wno_SAR}, or
$f_S$ and $f_W$ are not good order parameters for this model.  We suspect the
former; in every manner except this crossing issue, $f_S$ and $f_W$ behave as
expected.  (In the absence of clear crossing behavior for $f_S$ and $f_W$, for
the sake of discussion we nonetheless provide best visual estimates for
$\hat{T}_c(\alpha)$ for $f_S$ and $f_W$.  These will not be used for further
analysis toward our final result.)

\subsection{Verification of finite-size-scaling hypothesis}
\label{subsec:collapse}

\begin{figure}
\begin{center}
\psfragscanon
\includegraphics[scale=0.6]{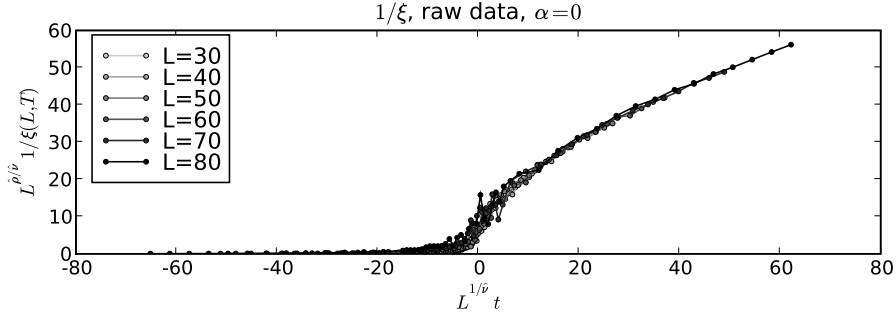}
\caption
	{Collapse plot for order parameter $1/\xi$.
	\label{fig:fS_collapse}}
\end{center}
\end{figure}

Now that we have estimated $\rho_S$, $\nu$, and $T_c(\alpha)$ for each of the
five order parameters $S$, we  may plot $L^{\rho_S/\nu} S_L(T,\alpha)$ as a
function of $L^{1/\nu}t$.  This is a plot of the scaling function $Q_S$.  If
the hypothesis is correct, the curves for all $L$ should coincide, or collapse,
to within sampling error --- which they do (e.g. figure \ref{fig:fS_collapse}).

\subsection{Determination of the shift in critical temperature}
\label{subsec:Delta_Tc_and_const}

\begin{figure}
\begin{center}
\psfragscanon
\includegraphics[scale=0.6]{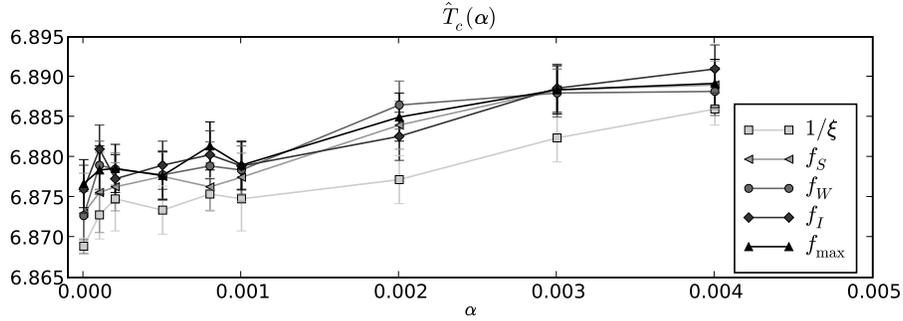}
\caption
	{Critical temperature as function of $\alpha$.
	\label{fig:Tc_alpha}}
\end{center}
\end{figure}
\begin{figure}
\begin{center}
\psfragscanon
\includegraphics[scale=0.6]{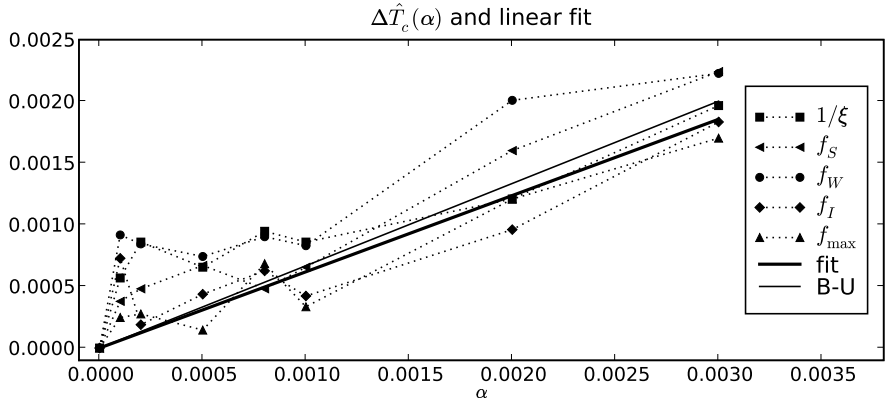}
\caption
	[Shift in critical temperature as function of $\alpha$.]
	{Shift in critical temperature, and linear fit, as function of $\alpha$.
	Recall from equation \eqref{eqn:Delta_Tc} that
	$\Delta T_c(\alpha) = \frac{T_c(\alpha)-T_c(0)}{T_c(0)}$.
	Order parameters $f_S$ and $f_W$ were omitted
	from the fit, due to lack of crossing behavior; $\alpha=0.004$ was
	omitted due to onset of curvature of $T_c(\alpha)$.
	The heavy solid line shows a linear fit with empirically determined
	constant of proportionality; the lighter solid line is the comparison value
	of Betz and Ueltschi (slope $2/3$) for decaying cycle weights and continuum
	point positions.
	\label{fig:Delta_Tc_cmp}}
\end{center}
\end{figure}

As discussed in section \ref{subsec:conj}, we are seeking a linear
relationship between $\Delta T_c(\alpha)$ and $\alpha$, with constant $c$.
This can be visualized in figure \ref{fig:Delta_Tc_cmp}, which is obtained from
the $T_{c,S}(\alpha)$ data of figure \ref{fig:Tc_alpha} using equation
\eqref{eqn:Delta_Tc}.  We start with all the $(\alpha,\Delta T_c(\alpha))$ data
points from section \ref{subsec:crossing}.  We omit values obtained using
$f_S$ and $f_W$, due to the aforemention lack of crossing behavior.  We also
omit values obtained using $\alpha=0.004$, since the critical-temperature plots
of figure \ref{fig:Tc_alpha} suggests that this starts to exceed the domain of
linear approximation.  We perform a linear regression with error bars
\cite{bib:young} on the $(\alpha,\Delta T_c(\alpha))$ data points.  We use a
slope-only fit, rather than a slope-intercept fit, since $\Delta T_c(\alpha)$
has zero intercept by its definition.  We find
\begin{align*}
	c &= 0.618 \pm 0.086 \textrm{  (2 $\sigma$ error bar)}.
\end{align*}
Within experimental uncertainty, this result, for points on the lattice with
Ewens cycle-weights, matches the $c$ value of equation \eqref{eqn:rho_c} for
point positions varying on the continuum with decaying-cycle-weight
interactions.

\subsection{Constancy of the macroscopic-cycle quotient}
\label{subsec:mcq_const}

\begin{figure}
\begin{center}
\psfragscanon
\includegraphics[scale=0.55]{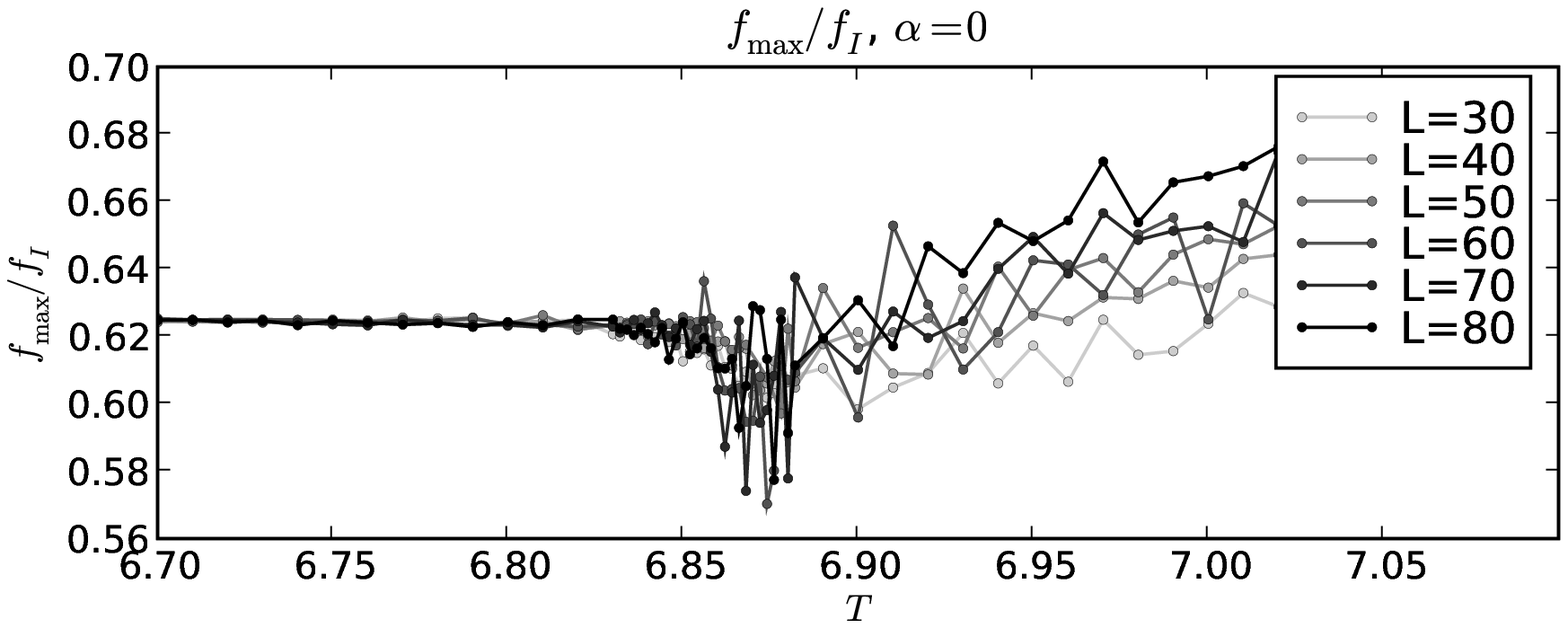}
\includegraphics[scale=0.55]{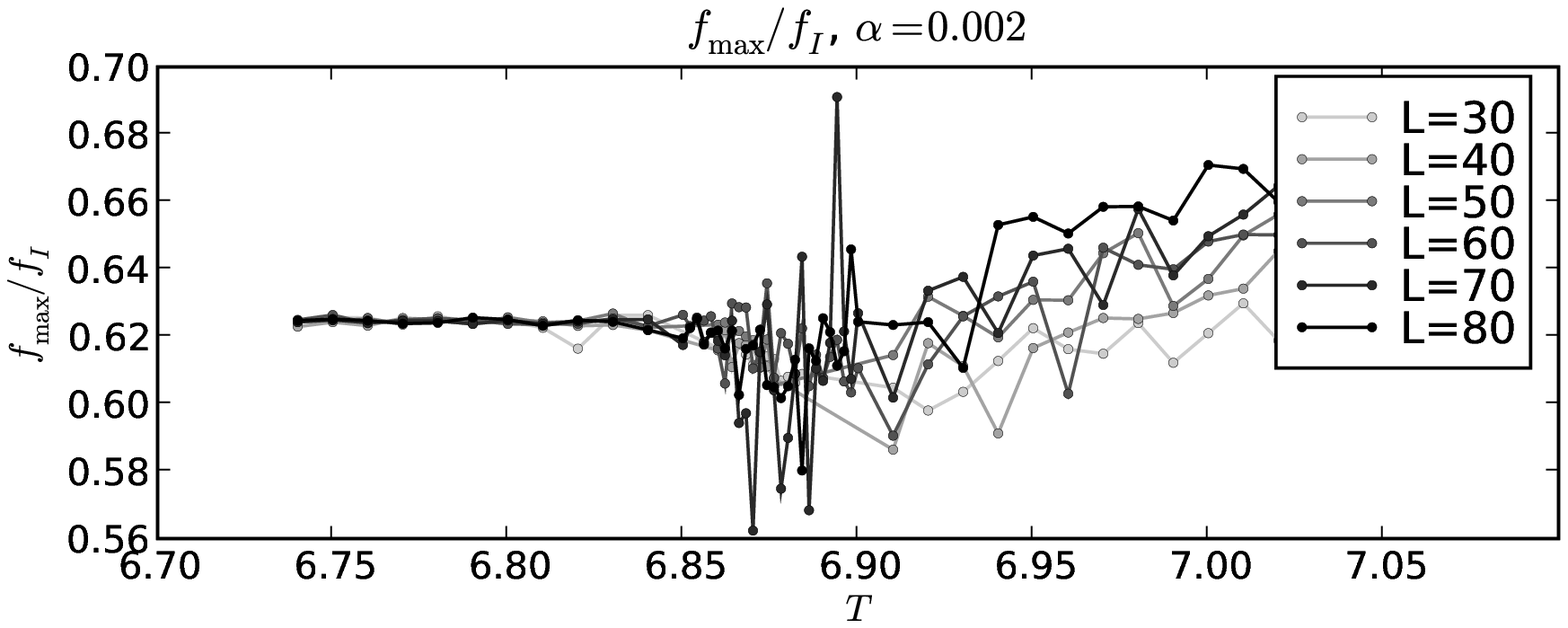}
\caption
	{Macroscopic-cycle quotient $f_M/f_I$ for $\alpha=0, 0.002$.
	\label{fig:fMfI_curves}}
\end{center}
\end{figure}

As discussed in section \ref{subsec:conj}, we hypothesize that the
macroscopic-cycle quotient $f_M / f_I$ in the infinite-volume limit is
dependent on $\alpha$ but is constant in $T$ where it is defined, i.e.  for
$T<T_c$ since $f_I=0$ for $T>T_c$.  This may be visualized by comparing figures
such as \ref{fig:E_ellmax_N}:  one sees that $f_M$ and $f_I$ appear to have the
same critical exponent.  Alternatively, one may plot the ratio $f_M/f_I$
(figure \ref{fig:fMfI_curves}).  In the infinite-volume limit, $f_I$ is zero
for $T>T_c$ and so we are interested only in the values of the quotient for
$T<T_c$.  In that region, the quotient does indeed appear to be constant in
$T$.

\begin{figure}
\begin{center}
\psfragscanon
\includegraphics[scale=0.50]{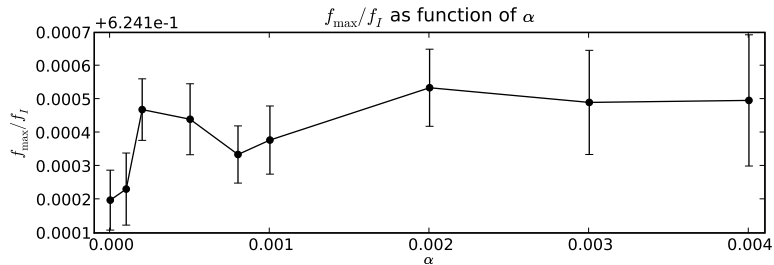}
\caption{$f_M/f_I$ as a function of $\alpha$.
	\label{fig:fMfI_alpha}}
\end{center}
\end{figure}

We test this constancy hypothesis as follows.  The respective critical
exponents are $\rho_M$ and $\rho_I$.  The estimators are $\rhohat_M$ and
$\rhohat_I$, computed by averaging over several different values of $L$ and
$\alpha$ as described in section \ref{subsec:rhohat_extrap}.  Treating these
estimators as normally distributed (as justified by the raw data), we obtain
the standard deviations of the $\rhohat_{M,I}(L,\alpha)$ samples, along with the
standard deviations of the means $\rhohat_{M,I}$:
\input{fMfIseparate.tex}
The difference $\rhohat_M - \rhohat_I$ is also normally distributed about the
true mean $\rho_M - \rho_I$, but $\rhohat_M$ and $\rhohat_I$ are not
independent since they are sample means of random variables computed from the
same Markov chain Monte Carlo sequence of permutations.  Thus
\begin{align*}
	\Var(\rhohat_M - \rhohat_I) &= \Var(\rhohat_M) + \Var(\rhohat_I)
		- 2 \Cov(\rhohat_M, \rhohat_I).
\end{align*}
Computing the sample covariance of the $\rhohat_M(L,\alpha)$ and
$\rhohat_I(L,\alpha)$ data series, we obtain the covariance and resulting
standard error $s_d$ of the difference
\input{fMfIcov.tex}
Normalizing, we find
\input{fMfInormalize.tex}
We hypothesize $\rho_M -  \rho_I = 0$; the estimated value $\rhohat_M -
\rhohat_I$ lies comfortably within a standard deviation of this.  We note,
moreover, that the value of $f_M/f_I$, while constant in $T$, trends upward
with $\alpha$ (see table \ref{table:fMfI_alpha} and figure
\ref{fig:fMfI_alpha}).  This merits further investigation.

\subsection{Conclusions}
\label{subsec:conclusions}


(1) For annealed point positions, equation \eqref{eqn:theo_Tc} gives $T_c(0)
\approx 6.625$.  Our result $T_c(0) = 6.873 \pm 0.006$ ($2\sigma$ error bar)
unambiguously shows that the lattice structure modifies the critical
temperature, even in the non-interacting ($\alpha=0$) case.


(2) As detailed in section \ref{subsec:Delta_Tc_and_const}, we find that the
reduced shift in critical temperature as a function of interaction parameter
$\alpha$ is
\begin{align*}
	\Delta T_c(\alpha) &\approx \frac{T_c(\alpha) - T_c(0)}{T_c(0)}
		= c \alpha
\end{align*}
with
\begin{align*}
	c &= 0.618 \pm 0.086 \quad\textrm{($2\sigma$ error bar)}.
\end{align*}
This is compatible (section \ref{subsec:conj}) with the related result of
\cite{bib:bu08}.  Even though the lattice structure changes the critical
temperature (conclusion 1), the \emph{shift} in critical temperature is
unaffected.


(3) As described in section \ref{subsec:conj}, Shepp and Lloyd
\cite{bib:shepplloyd} find that $\bbE[\ellmax]/N \approx 0.6243$ for
uniform-random (non-spatial) permutations.  For spatial permutations, we define
a macroscopic-cycle quotient $\bbE[\ellmax]/N f_I$ which is the ratio of mean
maximum cycle length as a fraction of the number of sites in long cycles.  Our
result (table \ref{table:fMfI_alpha}) is compatible with that of Shepp and
Lloyd for the non-interacting case, with an increase which appears to be linear
as a function of interaction parameter $\alpha$.  Our result is also compatible
with \cite{bib:gru}, which (among other conclusions) recovered the Shepp and
Lloyd result for the $\alpha=0$ case.

\section{Future work}
\label{sec:future_work}

Now that the $\alpha$-dependence of the macroscopic-cycle quotient's constant
upon $\alpha$ has been found empirically, one would next like to explain that
dependence analytically.

Ideally, one would have an algorithm to permit odd winding numbers, as
discussed in section \ref{subsec:wno_SAR}.

Sampling from the true Bose-gas distribution using the random-cycle model
requires three changes.  First, one needs to conduct simulations using the
Bose-gas interaction (equation \eqref{eqn:results_Bose_H}) rather than the
cycle-weight interaction (equation \eqref{eqn:results_RCM_H}).  The interaction
term $V$ is a CPU-intensive Brownian-bridge computation \cite{bib:bu07};
unpublished work of Ueltschi and Betz shows that it may be approximated in the
weak-interaction case by a simpler Riemann integral.  Precomputed tables and
interpolation may make use of this integral feasible.
Second, point positions must be allowed to vary on the continuum.  This entails
a second type of Metropolis step, in addition to that shown in section
\ref{subsec:SO_algorithm}.
Third, since points are no longer held fixed on the lattice, it is no longer
trivial to find nearest neighbors.  Software efficiency requires a hierarchical
partitioning of $\Lambda$.
The second and third points simply require a software effort.  Implementing
them will be worthwhile only if the interaction terms can be simplified to the
point that they are computationally feasible, which is a mathematical effort.

\section{Acknowledgements}
\label{sec:acknowledgements}

The author's doctoral dissertation work was done with co-advisors Daniel
Ueltschi and Tom Kennedy.  Daniel Gandolfo contributed several helpful
discussions.  The author was supported for three semesters by the National
Science Foundation, through NSF grant DMS-0601075 as well as the University of
Arizona Department of Mathematics VIGRE grant.  The author also wishes
to acknowledge the helpful comments provided by both of the anonymous
reviewers of the first version of this paper.



%% file: fMfIvalues.tex
\begin{center}
\begin{tabular}{|l|l|l|l|}
\hline $\alpha$ & Mean & Std.err. & Count \\
\hline
\hline  0.000 & 0.6242981 & 0.0000897 & 78 \\
\hline  0.0001 & 0.6243312 & 0.0001079 & 78 \\
\hline  0.0002 & 0.6245691 & 0.0000921 & 72 \\
\hline  0.0005 & 0.6245402 & 0.0001062 & 66 \\
\hline  0.0008 & 0.6244347 & 0.0000856 & 72 \\
\hline  0.001 & 0.6244779 & 0.0001020 & 60 \\
\hline  0.002 & 0.6246345 & 0.0001154 & 42 \\
\hline  0.003 & 0.6245906 & 0.0001559 & 48 \\
\hline  0.004 & 0.6245966 & 0.0001964 & 42 \\
\hline
\end{tabular}
\end{center}

%% file: rhoest.tex
\begin{center}
\begin{tabular}{|l|l|l|}
\hline $\nuhat$    & 0.5559 & $\pm$ 0.0037 \\
\hline $\rhohat_S$ & 0.6201 & $\pm$ 0.0065 \\
\hline $\rhohat_W$ & 0.7750 & $\pm$ 0.0073 \\
\hline $\rhohat_I$ & 0.7451 & $\pm$ 0.0052 \\
\hline $\rhohat_M$ & 0.7486 & $\pm$ 0.0059 \\
\hline
\end{tabular}
\end{center}

%% file: fMfIseparate.tex
\begin{align*}
  \rhohat_M        &= 0.7482   & \rhohat_I        &= 0.7445   \\
  s_M              &= 0.0428   & s_I              &= 0.0374   \\
  n_M              &= 50       & n_I              &= 50       \\
  s_M / \sqrt{n_M} &= 0.006059 & s_I / \sqrt{n_I} &= 0.005295.
\end{align*}

%% file: fMfIcov.tex
\begin{align*}
  \Cov(\rhohat_M, \rhohat_I) &= 0.0004 & s_d/\sqrt{n}   &= 0.0070.
\end{align*}

%% file: fMfInormalize.tex
\begin{align*}
  \rhohat_M - \rhohat_I &= 0.0037  & \frac{\rhohat_M - \rhohat_I}{s_d/\sqrt{n}} &= \frac{0.0037}{0.0070} = 0.5293.
\end{align*}

%% file: resultsbib.tex

%% file: rcm-ewens-results.bbl
\begin{thebibliography}{}

\bibitem[BBHLV]{bib:baymetal}
Baym,~G., Blaizot,~J.-P., Holzmann,~M., Lalo\"e,~F., and Vautherin,~D.
\newblock {\em Bose-Einstein transition in a dilute interacting gas}.
\newblock \texttt{arXiv:cond-mat/0107129v2}.
\newblock Eur. Phys. J. B 24, 107-124 (2001).

\bibitem[Berg]{bib:berg}
Berg,~B.
\newblock {\em Markov Chain Monte Carlo Simulations and Their Statistical
	Analysis}.
\newblock World Scientific Publishing (2004).

\bibitem[BPS06]{bib:bps06}
Boninsegni,~M., Prokof'ev,~N.V., and Svistunov,~B.V.
\newblock {\em Worm algorithm and diagrammatic Monte Carlo:  A new approach
	to continuous-space path integral Monte Carlo simulations.}
\newblock Physical Review E \textbf{74}, 036701 (2006).

\bibitem[BU07]{bib:bu07}
Betz,~V. and Ueltschi,~D.
\newblock{\em Spatial random permutations and infinite cycles}.
\newblock \texttt{arXiv:0711.1188}.
\newblock Commun. Math. Phys. 285, 469-501 (2009).

\bibitem[BU08]{bib:bu08}
Betz,~V. and Ueltschi,~D.
\newblock {\em Spatial random permutations with small cycle weights.}
\newblock \texttt{arXiv:0812.0569v1}.
\newblock Probabl. Th. Rel. Fields (2010).

\bibitem[CGGP]{bib:cggp}
Caracciolo,~S., Gambassi,~A., Gubinelli,~M., and Pelisetto,~A.
\newblock {\em Finite-Size Scaling in the Driven Lattice Gas.}
\newblock \texttt{arXiv:cond-mat/0312175}.
\newblock Journal of Statistical Physics, vol. 115, Nos. 1/2, April 2004.

\bibitem[Ewens]{bib:ewens}
Ewens,~W.J.
\newblock {\em The sampling theory of selectively neutral alleles}.
\newblock Theor. Popul. Biol. 3, 87-112 (1972).

\bibitem[Feynman]{bib:feynman}
Feynman,~R.P.
\newblock {\em Atomic Theory of the $\lambda$ Transition in Helium}.
\newblock The Physical Review, vol. 91, no. 6 (1953).

\bibitem[Golomb]{bib:golomb}
Golomb,~S.W.
\newblock {\em Random permutations}.
\newblock Bull. Ameer. Math. Soc. 70 (1964), 747.

\bibitem[GRU]{bib:gru}
Gandolfo,~D., Ruiz,~J., and Ueltschi,~D.
\newblock {\em On a model of random cycles}.
\newblock \texttt{arXiv:cond-mat/0703315}.
\newblock Statist. Phys. 129, 663-676 (2007).

\bibitem[Kerl]{bib:kerl}
Kerl,~J.
\newblock {\em Critical behavior for the model of random spatial permutations}.
\newblock Doctoral dissertation, University of Arizona, 2010.

\bibitem[LB]{bib:landaubinder}
Landau,~D.P. and Binder,~K.
\newblock {\em A Guide to Monte Carlo Simulations in Statistical Physics}
	(2nd ed.).
\newblock Cambridge University Press (2005).

\bibitem[NR]{bib:nr}
Press,~W. et al.
\newblock {\em Numerical Recipes} (2nd ed.).
\newblock Cambridge University Press (1992).

\bibitem[PC87]{bib:pc87}
Pollock,~E.L. and Ceperley,~D.M.
\newblock {\em Path-integral computation of superfluid densities.}
\newblock Physical Review B, vol. 36, no. 16 (1987).

\bibitem[PO]{bib:penons}
Penrose,~O. and Onsager,~L.
\newblock {\em Bose-Einstein Condensation and Liquid Helium}.
\newblock The Physical Review, vol. 104, no. 3 (1956).

\bibitem[PST98]{bib:pst98}
Prokof'ev,~N.V., Svistunov,~B.V., and Tupitsyn,~I.S.
\newblock {\em Exact, complete, and universal continuous-time worldline Monte
	Carlo approach to the statistics of discrete quantum systems.}
\newblock Journal of Experimental and Theoretical Physics, vol. 87, no. 2
	(1998).

\bibitem[SL]{bib:shepplloyd}
Shepp,~L.A. and Lloyd,~S.P. 
\newblock{\em Ordered Cycle Length in a Random Permutation}.
\newblock Trans. Amer. Math. Soc. 121, (1966), 340-357.

\bibitem[S\"ut\H{o}1]{bib:suto1}
S\"ut\H{o},~A.
\newblock {\em Percolation transition in the Bose gas.}
\newblock J. Phys. A: Math. Gen. \textbf{26} (1993) 4689-4710.

\bibitem[S\"ut\H{o}2]{bib:suto2}
S\"ut\H{o},~A.
\newblock {\em Percolation transition in the Bose gas II.}
\newblock J. Phys. A: Math. Gen. \textbf{35} (2002) 6995-7002.

\bibitem[SU09]{bib:su09}
Seiringer,~R. and Ueltschi,~D.
\newblock {\em Rigorous upper bound on the critical temperature of dilute Bose
gases}.
\newblock \texttt{arXiv.org:0904.0050}.
\newblock Phys. Rev. B 80, 014502 (2009).

\bibitem[U07]{bib:qmath}
Ueltschi,~D.
\newblock {\em The model of interacting spatial permutations and its relation
to the Bose gas.}
\newblock \texttt{arXiv:0712.2443v3}.
\newblock Mathematical Results in Quantum Mechanics, pp. 225-272,
World Scientific (2008).

\bibitem[Young]{bib:young}
Young,~H.D.
\newblock {\em Statistical Treatment of Experimental Data}.
\newblock McGraw-Hill (1962).

\end{thebibliography}
